\documentclass[%
floatfix,
preprint,
nobibnotes,
 amsmath,amssymb,
 aps,
 pra,
floatfix,
]{revtex4-1}

\usepackage{graphicx}% Include figure files
\usepackage{dcolumn}% Align table columns on decimal point
\usepackage{bm}% bold math
\usepackage{hyperref}% add hypertext capabilities
\usepackage[mathlines]{lineno}% Enable numbering of text and display math
\begin{document}

\title{On the calculation of second-order magnetic properties using subsystem approaches in the relativistic framework}

\author{Ma\l gorzata Olejniczak}
 \email{gosia.olejniczak@univ-lille1.fr}
\affiliation{Universit\'e de Lille, CNRS, UMR 8523 -- PhLAM -- Physique des Lasers, Atomes et Mol\'ecules, F-59000 Lille, France Fax: +33-3-2033-7020; Tel: +33-3-2043-4163}

\author{Radovan Bast}
 \email{radovan.bast@uit.no}
\affiliation{High Performance Computing Group, UiT The Arctic University of Norway, N--9037 Troms\oe, Norway. Tel: +47-776-44117}

\author{Andr\'e Severo Pereira Gomes}
 \email{andre.gomes@univ-lille1.fr (corresponding author)}
\affiliation{Universit\'e de Lille, CNRS, UMR 8523 -- PhLAM -- Physique des Lasers, Atomes et Mol\'ecules, F-59000 Lille, France Fax: +33-3-2033-7020; Tel: +33-3-2043-4163}

\date{\today}% It is always \today, today,

\begin{abstract}
We report an implementation of the nuclear magnetic resonance (NMR) shielding
($\sigma$), isotope-independent indirect spin-spin coupling ($K$) and the
magnetizability ($\xi$) tensors in the frozen density embedding scheme
using the four-component (4c) relativistic Dirac--Coulomb (DC) Hamiltonian and
the non-collinear spin density functional theory. The formalism takes
into account the magnetic balance between the large and the small components of
molecular spinors and assures the gauge-origin independence of NMR shielding
and magnetizability results. This implementation has been applied to
hydrogen-bonded HXH$\cdots$OH$_2$ complexes (X = Se, Te, Po) and compared with
the supermolecular calculations and with the approach based on the integration
of the magnetically induced current density vector. A comparison with the
approximate Zeroth-Order Regular Approximation (ZORA) Hamiltonian indicates
non-negligible differences in $\sigma$ and $K$ in the HPoH$\cdots$OH$_2$
complex, and calls for a thorough comparison of ZORA and DC in the description
of environment effects on NMR parameters for molecular systems with heavy
elements. (\textbf{Electronic Supplementary Information available at \url{http://dx.doi.org/10.5281/zenodo.179720}})
\end{abstract}

\maketitle

\section{Introduction}
\label{sec:intro}

The response to magnetic fields can be of great help in investigating molecular
systems in complex environments. This is perhaps best illustrated by the
widespread use of experimental techniques such as NMR spectroscopy to
characterize compounds in condensed phase, including disordered and amorphous
solids. The great sensitivity of this technique to any changes in electronic
structure in the vicinity of a probed nucleus, triggered for instance by
specific inter- and intramolecular interactions (e.g.\ hydrogen bonds) found in
complex biological systems (e.g. proteins, peptides, amino
acids),\cite{oldfield-annurevphyschem-53-349-2002,frank-jctc-8-1480-2012}
catalysts,\cite{nmr_catalysis} paramagnetic
systems\cite{pintacuda-modern-nmr-book,autschbach-arcc-11-3-2015} and
radioactive compounds containing actinide
atoms\cite{yasuoka-science-336-901-2012,martel-revsciinstrum-84-055112-2013} is
often unrivaled by other experimental techniques.

Due to the complexity of such systems, theoretical tools can be of extreme
importance to aid experimentalists interpret their results. Most theoretical
approaches hinge on the recognition that, in the weak magnetic field regime
under which most experiments are carried out, the magnetic field can be treated
as a perturbation, the energy of a molecule can be Taylor-expanded in terms of
perturbation strengths and the effect of a perturbation on a given system can
be evaluated via response theory.\cite{helgaker-cr-112-543-2012}

For closed-shell systems, a static magnetic field induces only even-order
changes in the total energy:\cite{helgaker-chemrev-99-293-1999}
  \begin{equation}
    E(\mathbf{\varepsilon}) = E_0 + \frac{1}{2} \frac{d^2E}{d\varepsilon_1 d\varepsilon_2} \varepsilon_1 \varepsilon_2 +  \frac{1}{4!} \frac{d^4E}{d\varepsilon_1\ldots d\varepsilon_4} \varepsilon_1\ldots \varepsilon_4 + \ldots,
  \end{equation}
where $E_0$ denotes the energy at zero field and $\{\varepsilon_n\}$ the field
strengths of applied perturbations collected in vector $\mathbf{\varepsilon}$.
The coefficients of this expansion, taken in the zero-field limit, define
molecular properties in the Born--Oppenheimer approximation. In this paper we
focus on three second-order magnetic properties arising from a perturbation of
an external field $\vec{B}$ or the field of nuclear magnetic dipole moments,
$\{\vec{m}_A\}$: the NMR shielding tensor of a nucleus K,
  \begin{equation}
    \sigma^K_{\alpha\beta} =  \frac{d^2E}{dB_\alpha dm_{K;\beta}} \biggr|_{\vec{B},\{\vec{m}_A\} = 0},
    \label{eq:nmr}
  \end{equation}
the reduced spin-spin coupling tensor of nuclei K and L,
    \begin{equation}
      K^{KL}_{\alpha\beta} =  \frac{d^2E}{dm_{K;\alpha}dm_{L;\beta}} \biggr|_{\{\vec{m}_A\} = 0},
      \label{eq:spin-spin}
    \end{equation}
related to the indirect spin-spin coupling constants observed in NMR experiment,
$J^{KL} = (\hbar/2\pi)\gamma_\text{K}\gamma_\text{L} K^{KL}_{\alpha\beta}$, with $\gamma_\text{M}$ denoting the
gyromagnetic ratio of a given isotope of $M$, and the molecular magnetizability tensor,
  \begin{equation}
    \xi_{\alpha\beta} =  - \frac{d^2E}{dB_\alpha dB_\beta} \biggr|_{\vec{B} = 0},
    \label{eq:magnetizability}
  \end{equation}
which, unlike the first two, is not a local but rather an extensive property as
it does not depend on the magnetic moment of a given nucleus -- though it may
play an important role in NMR spectroscopy by inducing changes in the local
magnetic fields which, in turn, will affect chemical
shifts.\cite{mcconnell-jcp-27-226-1957,sitkoff-pnmrs-32-165-1998,facelli-pnmrs-58-176-2011}

NMR properties can nowadays be routinely calculated with density functional
theory (DFT) for relatively large closed-shell systems and with methods based
on wave function theory (WFT) for much smaller
systems.\cite{jackowski-jaszunski-book-2016,nmr-book-kaupp,helgaker-chemrev-99-293-1999}
Calculating the magnetizability tensor is computationally more demanding due to
the slow convergence with the basis set size. This can be partly alleviated by
the use of London atomic orbitals
(LAOs),\cite{london-jpr-8-397-1937,helgaker-chemrev-99-293-1999} which also
remedy the gauge-origin dependence problem for properties arising from an
external magnetic field perturbation ($\xi, \sigma^K$), at the cost of more
complex derivations and more involved implementation.

Furthermore, it is now recognized that relativistic
effects\cite{autschbach-ptrsa-372-20120489-2014,repisky-jackowski-jaszunski-book-2016}
can be appreciable for
magnetizabilities\cite{yoshizawa-jcc-30-2550-2009,schwerdtfeger-jcp-134-204102-2011}
and are essential for a proper description of NMR properties, including for
those light elements neighbouring the heavy one(s) due to the so-called
"heavy-atom on the light atom" (HALA) effect.\cite{pyykko-mp-61-196-1987}.  NMR
properties obtained with approximate Hamiltonians such as the quasirelativistic
two-component (2c) ZORA Hamiltonian can differ significantly from those
obtained with a more rigorous treatment afforded by four-component
approaches as in the Dirac-Coulomb Hamiltonian, even for relatively light
elements such as those in the fourth row of the periodic table.  The
conventional wisdom has been that approximate 2c approaches, which are
computationally much cheaper than 4c ones, can be nevertheless reliable for
relative quantities such as chemical shifts, due to error cancellation. Recent
studies\cite{autschbach-mp-111-2544-2013,vicha-pccp-17-24944-2015} paint a more
nuanced picture, and seem to indicate that there may be significant differences
between Hamiltonians for relative quantities as well, notably for heavier
elements.

One must also take into account the effect of the surroundings on the molecular
properties, something preferably done with embedding approaches due to the steep
increase in computation cost for the explicit inclusion of e.g.\ solvent
molecules. As methods representing the environment in an implicit manner such
as PCM\cite{miertus-cp-55-117-1981} or
COSMO\cite{klamt-jcspt-2-799-1993,pye-tca-101-396-1999,cossi-jcc-24-669-2003}
have difficulty describing specific interactions such as hydrogen bonds, one is
better served by approaches such as frozen-density embedding (FDE),
\cite{gomes-arc-108-222-2012, jacob-wcms-4-325-2014,
wesolowski-chemrev-115-5891-2015} in which the total system is partitioned into
subsystems whose interaction is calculated at DFT while the subsystems
themselves can be treated with DFT (DFT-in-DFT embedding) or WFT (WFT-in-DFT
embedding).  FDE has been applied to the calculation of molecular properties
arising from electric
perturbations,\cite{neugebauer-jcp-126-134116-2007,neugebauer-jcp-131-084104-2009,hofener-jcp-136-044104-2012,hofener-jcp-139-104106-2013,hofener-jcp-137-204120-2012}
and in particular employing 4c
Hamiltonians,\cite{gomes-pccp-10-5353-2008,gomes-pccp-15-15153-2013} though
there have been only a few studies of magnetic properties: NMR
shieldings\cite{jacob-jcp-125-194104-2006,bulo-jpca-112-2640-2008} and indirect
spin-spin couplings,\cite{gotz-jcp-140-104107-2014} the latter using the ZORA
Hamiltonian.

The aim of this paper is therefore to bridge this gap and propose an FDE
implementation that is capable of treating general second-order magnetic
properties with the 4c Dirac-Coulomb (DC) Hamiltonian, by extending the general
framework for response properties\cite{hofener-jcp-136-044104-2012} in line
with the 4c DFT simple magnetic balance (sMB)
framework.\cite{olejniczak-jcp-136-014108-2012} We also investigate the
real-space determination of NMR shielding via the integration of magnetically
induced currents and its use for understanding the effect of approximations
introduced in practical DFT-in-DFT calculations. This will allow for
investigating the suitability of approximate 2c approaches for the calculation
of chemical shifts and provide a way to incorporate environment effects in the
determination of shielding scales and the nuclear magnetic dipole
moments,\cite{antusek-cpl-660-127-2016} a field which has received renewed
interest in recent
years.\cite{lantto-jpca-115-10617-2011,komorovsky-jcp-142-091102-2015,malkin-jphyschemlett-4-459-2013,demissie-jcp-143-164311-2015}

The paper is organized as follows: in the next section we present an overview
of the theoretical formulation, followed by the presentation of
proof-of-principle calculations on the H$_2$X--H$_2$O (X = Se, Te, Po) family
of compounds, where we also compare the description of environment effects with
2c and 4c Hamiltonians for NMR shieldings and indirect spin-spin couplings.
Such a comparison for magnetizabilities is currently not possible as ours is,
to the best of our knowledge, the first FDE implementation.

\section{Theory}
\label{sec:theory}

We begin by briefly summarizing  the theory for NMR
shieldings,\cite{aucar-jcp-110-6208-1999,ilias-jcp-131-124119-2009,olejniczak-jcp-136-014108-2012}
magnetizability\cite{ilias-mp-111-1373-2013} and NMR spin-spin
couplings\cite{aucar-jcp-110-6208-1999,visscher-jcc-20-1262-1999,enevoldsen-jcp-112-3493-2000}
in closed-shell molecules with the 4c relativistic DC Hamiltonian and
mean-field methods and its implementation in the DIRAC\cite{DIRAC15} software,
followed by the general FDE framework for molecular
properties\cite{hofener-jcp-136-044104-2012} and its extension to magnetic
properties in relativistic framework.

Throughout the text, $i,j,\ldots$ denote occupied molecular orbitals, $a,
b,\ldots$ virtual orbitals and $p, q,\ldots$ orbitals in general. Greek indices
are used for the three Cartesian components and Latin indices are used for the
components of four-component vector. The summation over repeated indices is
assumed. The SI-based atomic units are employed (~$\hbar = m_e = e =
1/(4\pi\epsilon_0) = 1$).\cite{whiffen-pac-50-75-1978} As we restrict ourselves
to closed-shell systems represented by a single Slater determinant, we employ
the following parametrization in second quantization for the unperturbed
wavefunctions:
  \begin{equation}
    |\tilde{0}\rangle = \exp \left(-\hat{\kappa}\right)|0 \rangle;
    \qquad
    \hat{\kappa} = \kappa_{ai} \hat{a}^\dagger \hat{i} - \kappa_{ai}^* \hat{i}^\dagger \hat{a},
    \label{eq:kappa}
  \end{equation}
where $\hat{\kappa}$ is an anti-Hermitian operator represented by a matrix of
orbital rotation amplitudes, which serve as variational parameters in
optimization of the ground state and its response to an external perturbation.

\subsection{Molecules in magnetic fields in a 4c framework}
\label{subsec:mag_general}

\paragraph*{Molecular Hamiltonian}
Starting with the generic form of the Hamiltonian in the Born-Oppenheimer approximation,
  \begin{equation}
    \hat{H} = \sum_i \hat{h}_i + \sum_{i < j} \hat{g}_{ij} + V_{NN},
    \label{eq:general_hamiltonian}
  \end{equation}
with $V_{NN}$ denoting a classical repulsion potential of clamped nuclei, it is
the choice of one- and two-electron operators - $\hat{h}_i$ and $\hat{g}_{ij}$,
that determines whether and which relativistic effects are
included.\cite{saue-cpc-12-3077-2011} The one-electron part corresponds to the
Dirac operator, which in the presence of a uniform external magnetic field
($\vec{B}$) and magnetic dipole moments of nuclei ($\{\vec{m}_K\}$) reads:
  \begin{equation}
    \hat{h} = \hat{h}_0 + \vec{B} \cdot \hat{h}_B + \sum_K \vec{m}_K \cdot \hat{h}_{m_K},
    \qquad
    \hat{h}_0 = \beta'mc^2 + c(\vec{\alpha}\cdot\vec{p}) + v_\text{nuc},
    \label{eq:1el_hamiltonian_mag}
  \end{equation}
where $\alpha$ and $\beta' = \beta - 1_{4\times4}$ are 4$\times$4 Dirac
matrices in their standard representation, $c$ is the speed of
light.\cite{dyall-faegri-book-2007} The Zeeman operator, $\hat{h}_B$, and the
hyperfine operators, $\hat{h}_{m_K}$, are defined as:
  \begin{equation}
    \hat{h}_B = \frac{1}{2} (\vec{r}_G \times c\vec{\alpha}),
    \qquad
    \hat{h}_{m_K} = \frac{1}{c^2}\frac{\vec{r}_K \times c \vec{\alpha}}{r_K^3},
    \label{eq:Zeeman_hyperfine}
  \end{equation}
where $\vec{r}_X = \vec{r} - \vec{R}_X$ with an arbitrary gauge origin
$\vec{R}_G$ and the center of nucleus $K$ in $\vec{R}_K$.  The two-electron
part of the 4c DC Hamiltonian is restricted to the Coulomb potential, which in
relativistic regime also includes spin-same-orbit
interaction.\cite{saue-cpc-12-3077-2011}

\paragraph*{One-electron basis} The 4c molecular spinors, eigenfunctions of the
DC Hamiltonian, are expanded in scalar finite basis
sets.\cite{dyall-faegri-book-2007} The small component basis functions
($\chi^\text{S}$) are generated from the large component functions
($\chi^\text{L}$) by the restricted kinetic balance (RKB) or
the restricted magnetic balance (RMB)
prescription, which properly describes the relation between large and small 
components in the presence of external magnetic fields. The question of magnetic balance 
in 4c calculations was initially addressed by Aucar and coworkers~\cite{aucar-jcp-110-6208-1999} and 
Kutzelnigg,\cite{kutzelnigg-jcc-20-1199-1999,kutzelnigg-pra-67-032109-2003} though
later investigations have shown these yielded mixed results.\cite{visscher-aqc-48-369-2005}
Later, Komorovsky and coworkers,\cite{komorovsky-jcp-128-104101-2008,repisky-cp-356-236-2009,
komorovsky-jcp-132-154101-2010} followed by Cheng\cite{cheng-jcp-131-244113-2009} and Reynolds,\cite{reynolds-pccp-17-14280-2015}
have revisited the question formally -- and computationally -- establishing RMB.

For properties explicitly dependent on an external magnetic field ($\sigma,
\xi$), the basis functions are replaced by London atomic orbitals,
  \begin{equation}
    \omega_\mu^K (\vec{r}) = \exp \left\{ -\frac{\text{i}}{2} \vec{B} \times (\vec{R}_K - \vec{R}_G) \cdot \vec{r} \right\}\chi_\mu^K (\vec{r}),
    \label{eq:LAO}
  \end{equation}
which guarantee the gauge-origin invariance of results in a finite basis
approximation. LAOs are also appealing when used in combination with RMB as
they make the magnetic balance atomic and easy to handle by the simple scheme
(sMB).\cite{olejniczak-jcp-136-014108-2012} The orthogonality of molecular
orbitals at all field strengths is ensured by \emph{connection matrices},
$T$,\cite{olsen-tca-90-421-1995} which couple the unmodified molecular orbitals
(UMOs) to the orthonormalized set (OMOs, $\{\breve{\psi}\}$),
  \begin{equation}
     \psi^\text{UMO}_q(\vec{B}) = \omega_{\mu} (\vec{B}) c_{\mu q} (0),
     \qquad
     \breve{\psi}_p(\vec{B}) = \psi^\text{UMO}_q(\vec{B}) T_{qp}(\vec{B}),
     \label{eq:psi_omo_umo}
  \end{equation}
yet for the price of more complex equations, as the wave function is now
dependent on a perturbation through $\omega_{\mu}(\vec{B})$ and $T(\vec{B})$ in
addition to $\kappa_{pq}(\vec{B})$.

\paragraph*{Spin Density Functional Theory} The choice of the spin-density
functional theory
(SDFT)\cite{barth-jpc-5-1629-1972,rajagopal-prb-7-1912-1973,jacob-ijqc-112-3661-2012}
is a compromise between a
desirable\cite{vignale-prb-37-10685-1988,vignale-prb-37-2502-1988,vignale-prl-59-2360-1987}
yet so far
unattainable\cite{lee-jcp-103-10095-1995,lee-cpl-229-225-1994,tellgren-jcp-140-034101-2014,furness-jctc-11-4169-2015}
DFT formalism for molecules in magnetic fields involving the current density,
and the conventional charge-density-only approaches whose density functional
approximations (DFAs) are developed to reproduce energy in the absence of
magnetic
perturbations.\cite{lutnaes-jcp-131-144104-2009,teale-jcp-138-024111-2013}
Also, due to the complexity of relativistic generalization of DFT
method,\cite{rajagopal-prb-7-1912-1973,relativistic-schwerdtfeger-ch10,dftbook-engel-dreizler-2011}
non-relativistic functionals are used with relativistic densities.

In this work, the non-collinear SDFT is employed, with the spin density
(calculated as a norm of spin magnetization vector) and the charge density as
basic variables, expressed together as a general density
component:\cite{bast-ijqc-109-2091-2009,komorovsky-jcp-128-104101-2008,olejniczak-jcp-136-014108-2012}
  \begin{equation}
    \rho_k (\vec{r}, \vec{B}) = \breve{\Omega}_{pq;k} (\vec{r}, \vec{B}) \tilde{D}_{pq}(\kappa)
    \qquad
    k \in \{0, x, y, z\},
    \label{eq:rho_k}
  \end{equation}
with the elements of the density matrix, $\tilde{D}_{pq} = \langle
\tilde{0}|p^\dagger q |\tilde{0} \rangle$ and the generalized overlap
distribution, $\breve{\Omega}$, expressed in OMO basis whenever LAOs are used:
  \begin{equation}
     \breve{\Omega}_{pq;k} = \breve{\psi}_{p}^{\dagger} \Sigma_{k} \breve{\psi}_{q};
     \qquad
     \Sigma_{0} = I_{4\times4},
     \quad
     \Sigma_{\mu} = \left[
        \begin{array}{cc}
        \sigma_{\mu}   & 0_{2\times2} \\
        0_{2\times2}   & \sigma_{\mu}
        \end{array}
      \right].
      \label{eq:Omega_k}
  \end{equation}
The ground state energy and the optimal $\rho_k$ are obtained by minimization
of the energy functional, $E[\rho_k]$, which can be written in a Kohn--Sham
(KS) manner as a sum of five terms:
  \begin{align}
    E[\rho_k] &= T_{s}[\rho_k] + E_\text{xc} [\rho_k] + V_{NN} \label{eq:Esdft_2} \\
              &+ \int \left( \rho_0 v_\text{nuc} + \sum_{\mu = x, y, z} \rho_\mu \cdot B_\mu \right) d\vec{r}
               + \frac{1}{2}\iint \frac{\rho_0(\vec{r}_1)\rho_0(\vec{r}_2)}{|\vec{r}_1   - \vec{r}_2|} d\vec{r}_1 d\vec{r}_2,
               \nonumber
  \end{align}
where $T_{s}$ is a kinetic energy of non-interacting electrons, $E_\text{xc}$ is the
exchange--correlation (xc) contribution and two last terms describe the
interaction of electrons with an electromagnetic potential and with other
electrons, respectively.  The minimization of Eq.~\ref{eq:Esdft_2} with respect
to $\rho_k$ (in the zero magnetic-field limit) yields the 4c KS equations for
the DC Hamiltonian:
  \begin{equation}
    \left( \beta'mc^2 + c(\vec{\alpha}\cdot\vec{p}) + v_\text{KS}[\rho_k]  \right) \psi = \varepsilon \psi,
    \label{eq:KS}
  \end{equation}
with the effective KS potential :
  \begin{equation}
    v_\text{KS} [\rho_k] = - \left( v_\text{nuc} + \int \frac{\rho_0(\vec{r}')}{|\vec{r} - \vec{r}'|} d\vec{r}' + \frac{\delta E_\text{xc}}{\delta \rho_k} \right).
    \label{eq:KSpot_s}
  \end{equation}

\paragraph*{Linear response (LR) at the 4c SDFT level} Considering now the case
of time-independent perturbations with strengths $\varepsilon_1$ and
$\varepsilon_2$, the second-order molecular property can be written as:
  \begin{equation}
    \left. \frac{d^2E}{d\varepsilon_1 d\varepsilon_2} \right|_{\varepsilon = 0}
    =
    \left. \frac{\partial^2E}{\partial \kappa_{pq} \partial \varepsilon_2} \frac{\partial \kappa_{pq}}{\partial \varepsilon_1} \right|_{\varepsilon=0}
    +
    \left. \frac{\partial^2E}{\partial \varepsilon_1 \partial \varepsilon_2}  \right|_{\varepsilon=0},
    \label{eq:d2E_dYdY}
  \end{equation}
assuming that the energy is optimized with respect to variational parameters at
all field strengths, $\partial E/\partial \kappa = 0$.  The first
contribution is determined perturbatively, with the first-order orbital
rotation amplitudes, $\partial \kappa / \partial \varepsilon$, obtained
from the LR equations:
  \begin{equation}
    0 = \left. \frac{d}{d\varepsilon_1} \left( \frac{\partial E}{\partial \kappa_{pq}} \right) \right|_{\varepsilon = 0} = \left. \left( \frac{\partial^2E}{\partial \kappa_{pq} \partial \varepsilon_1} + \frac{\partial^2E}{\partial \kappa_{pq} \partial \kappa_{rs}} \frac{\partial \kappa_{rs}}{\partial \varepsilon_1}  \right) \right|_{\varepsilon = 0},
    \label{eq:lr_long}
  \end{equation}
which can be recast in a matrix form as:\cite{saue-jcp-118-522-2003}
  \begin{equation}
    0 = \mathbf{E}_{\varepsilon_1}^{[1]} +  \mathbf{E}_{0}^{[2]} \mathbf{X}_{\varepsilon_1}.
     \label{eq:lr}
  \end{equation}
Here, $\mathbf{E}_{0}^{[2]}$ is the electronic Hessian,
$\mathbf{E}_{\varepsilon_1}^{[1]}$ the property gradient and
$\mathbf{X}_{\varepsilon_1}$ the solution vector yielding
$\{\kappa_{rs}^{\varepsilon_1}\}$.  While the Hessian is independent on a
perturbation, the property gradient is calculated as the first-order derivative
of the KS matrix with respect to field strength of applied perturbation,
  \begin{equation}
     \mathbf{E}_{\varepsilon_1}^{[1]} = \begin{bmatrix}
                  g^{\varepsilon_1}\\
                  g^{* \varepsilon_1}
                 \end{bmatrix},
     \qquad
     g_{ai}^{\varepsilon_1} = \left. \frac{\partial E_{\varepsilon_1}}{\partial \kappa_{ai}^*} \right|_{0} = \langle 0 |[-\hat{a}_i^\dagger \hat{a}_a, \hat{h}_{\varepsilon_1}] |0 \rangle =  -\tilde{F}_{ai}^{\varepsilon_1}.
     \label{eq:prop_grad}
  \end{equation}
In particular, if $\varepsilon_1 = \vec{B}$, the property gradient is
calculated in OMO basis and requires additional contributions involving
derivatives of LAOs and of matrices
T.\cite{ilias-jcp-131-124119-2009,olejniczak-jcp-136-014108-2012} Once
$\mathbf{X}_{\varepsilon_1}$ has been determined, one can construct the static
linear response function:
  \begin{equation}
    \langle \langle \varepsilon_1; \varepsilon_2 \rangle \rangle_0 = E_{\varepsilon_1}^{[1]\dagger} X_{\varepsilon_2} = -E_{\varepsilon_1}^{[1]\dagger} \left( E_0^{[2]} \right)^{-1} E_{\varepsilon_2}^{[1]},
    \label{eq:linresfun}
  \end{equation}
which constitutes the response contribution to the molecular property expressed
by the first term of Eq.~\ref{eq:d2E_dYdY}.  The second term of
Eq.~\ref{eq:d2E_dYdY} can be thought of as an expectation value, which due to
the linearity of the DC Hamiltonian in applied perturbations
(Eq.~\ref{eq:1el_hamiltonian_mag}) is non-zero only in perturbation-dependent
basis sets.

This brings about to the final form of the properties of interest in this
paper:
\begin{eqnarray}
K^{KL}_{\alpha\beta} &=&  \langle \langle m_{K;\alpha}; m_{L;\beta} \rangle \rangle_0 \label{eq:spinspin_simple} \\
\sigma^K_{\alpha\beta} &=&  \langle \langle m_{K;\alpha}; B_\beta \rangle \rangle_0 \label{eq:shield_simple} \\
\xi_{\alpha\beta} &=& -\left( \langle \langle B_\alpha; B_\beta \rangle \rangle_0 + \frac{\partial^2E}{\partial B_\alpha \partial B_\beta} \biggr|_0 \right) \label{eq:magnetizability_simple}
\end{eqnarray}
with the LAO basis used for the last two.

\subsection{Frozen Density Embedding}
\label{subsec:fde}
In FDE the total system is partitioned into interacting subsystems (for
simplicity here we shall consider only two, the one of interest (I) and the
other (II) representing the environment) implying a partition of the total
density and energy.\cite{gomes-arc-108-222-2012} We can further consider the
case of spin-density
FDE\cite{wesolowski-radft-371-2002,solovyeva-jcp-136-194104-2012} and partition
the generalized density component and the energy,
\begin{equation}
  \rho_k^\text{tot}(\vec{r}) = \rho_k^I(\vec{r}) + \rho_k^{II}(\vec{r})
  \label{eq:rhotot}
\end{equation}
  \begin{equation}
    E_\text{tot}[\rho_k^\text{tot}] = E_I[\rho_k^I] + E_{II}[\rho_k^{II}] + E_\text{int}[\rho_k^I, \rho_k^{II}]
    \label{eq:Etot},
  \end{equation}
where $E_{M}[\rho_k^M]$ is the energy of an isolated subsystem ($M = I, II$)
calculated from Eq.~\ref{eq:Esdft_2} and $E_\text{int}$ is the interaction energy
dependent on densities of both subsystems,
  \begin{align}
    E_\text{int}[\rho^{I}_k, \rho^{II}_k]
      &= E_\text{tot}[\rho_k^\text{tot}] - E_I[\rho_k^I] - E_{II}[\rho_k^{II}] \nonumber \\
      &=  \int \left[ \rho_0^{I}(\vec{r}) v_\text{nuc}^{II}(\vec{r}) + \rho_0^{II}(\vec{r}) v_\text{nuc}^{I}(\vec{r}) \right]d\vec{r}  + E_\text{nuc}^{I,II} \nonumber \\
      &+ \int \int \frac{\rho_0^{I}(\vec{r}_1) \rho_0^{II}(\vec{r}_2)}{|\vec{r}_1 - \vec{r}_2|} d\vec{r}_1d\vec{r}_2
      + E_\text{xc}^\text{nadd} + T_{s}^\text{nadd}.
      \label{eq:Eint}
  \end{align}
where $E_\text{nuc}^{I,II}$ is the nuclear repulsion energy between subsystems,
$E_\text{xc}^\text{nadd}$ and $T_s^\text{nadd}$ are the non-additive contributions defined
as:\cite{hofener-jcp-136-044104-2012}
  \begin{equation}
    X^\text{nadd} \equiv X^\text{nadd}[\rho_k^{I}, \rho_k^{II}] = X[\rho_k^\text{tot}] - X[\rho_k^{I}] - X[\rho_k^{II}]
    \label{eq:nonadd}.
  \end{equation}

In order to determine $\rho_k^{I}$ in the presence of other subsystem(s) with a
given generalized density $\rho_k^{II}$ one solves the 4c KS equations for a
constrained electron density (KSCED)\cite{wesolowski-jpc-97-8050-1993} which,
in the limit of zero magnetic field have the form
  \begin{equation}
    \left( \beta' mc^2 + c(\vec{\alpha} \cdot \vec{p}) + v_\text{KS}[\rho_k^I] + v_\text{emb}^I[\rho_k^{I}, \rho_k^{II}] \right) \psi^I(\vec{r}) = \varepsilon^I \psi^I(\vec{r})
    \label{eq:KSCED},
  \end{equation}
where an effective KS potential of Eq.~\ref{eq:KSpot_s} is augmented by the embedding potential,
  \begin{equation}
    v_\text{emb;k}^I(\vec{r}) = \frac{\delta E_\text{int}}{\delta \rho_k^I(\vec{r})}
                   = \frac{\delta E_\text{xc}^\text{nadd}}{\delta \rho_k^I} +  \frac{\delta T_{s}^\text{nadd}}{\delta \rho_k^I}  + v_\text{nuc}^{II}(\vec{r}) + \int \frac{\rho_0^{II}(\vec{r}')}{|\vec{r} - \vec{r}'|}d\vec{r}'
    \label{eq:embpot},
  \end{equation}
representing the interaction of subsystem $I$ with other subsystem(s). One can
also relax the constraints on $\rho_k^{II}$ by interchanging it with $\rho_k^I$
and solving the analogous KSCED equations in an iterative manner in the
so-called \emph{freeze-thaw}\cite{jacob-jcc-29-1011-2008} procedure.

FDE is formally exact in the limit of exact functionals describing the
non-additive exchange-correlation and kinetic energies, but for computational efficiency both are
generally obtained with approximate density functionals and grouped into a
single term,
\begin{equation}
E^\text{nadd}_\text{xck}[\rho_k^{I},\rho_k^{II}] = E^\text{nadd}_\text{xc}[\rho_k^{I},\rho_k^{II}] + T^\text{nadd}_s[\rho_k^{I},\rho_k^{II}].
\end{equation}
It should be noted that the currently available kinetic energy functionals have
a limited accuracy,\cite{gomes-arc-108-222-2012,jacob-wcms-4-325-2014} and
while sufficient for relatively weak interactions (eg. hydrogen
bonds),\cite{schluns-pccp-17-14323-2015} practical difficulties may emerge for
stronger ones requiring to replace the kinetic energy density functionals by
other approaches.\cite{fux-jcp-132-164101-2010,artiukhin-jcp-142-234101-2015}

\subsubsection{FDE molecular properties}

In what follows we shall discuss the contributions to second-order molecular
properties presented in Eq.~\ref{eq:d2E_dYdY} in a subsystem manner.  We use
separate sets of externally orthogonal orbitals for subsystems $I$ and
$II$,\cite{hofener-jcp-136-044104-2012,unsleber-pccp-2016} implying the
separate sets of orbital rotation coefficients,
$\kappa_{p_Mq_N} = \delta_{MN} \kappa_{p_Mq_M}$ for $M, N \in \{I, II\}$, and the parametrization of the total density (Eq.~\ref{eq:rhotot}):
\begin{equation}
  \rho_k^\text{tot}(\vec{r}, \kappa^I, \kappa^{II}) = \rho_k^{I}(\vec{r}, \kappa^{I})
                        + \rho_k^{II}(\vec{r}, \kappa^{II}),
  \label{eq:rhotot_parametrization}
\end{equation}
with $\kappa^M = \{ \kappa_{p_M q_M} \}$ for $M \in \{I, II\}$.

\paragraph{Linear response functions}
The electronic Hessian and property gradient are now subdivided into isolated
subsystem and interaction contributions\cite{hofener-jcp-136-044104-2012}
\begin{eqnarray}
\mathbf{E}_0^{[2]}
  &=& \begin{bmatrix}
        \mathbf{E}_{0}^{[2];M,M} & \mathbf{0} \\
        \mathbf{0} & \mathbf{E}_{0}^{[2];N,N}
      \end{bmatrix}
    + \begin{bmatrix}
        \mathbf{E}_{0,\text{int}}^{[2];M,M} & \mathbf{E}_{0;\text{int}}^{[2];M,N} \\
        \mathbf{E}_{0,\text{int}}^{[2];N,M} & \mathbf{E}_{0;\text{int}}^{[2];N,N}
    \end{bmatrix}, \label{eq:FDEhessian} \\
\mathbf{E}_{\varepsilon_1}^{[1]}
   &=& \begin{bmatrix}
         \mathbf{E}_{\varepsilon_1}^{[1];M} & \mathbf{E}_{\varepsilon_1}^{[1];N}
       \end{bmatrix}^\dagger
     + \begin{bmatrix}
         \mathbf{E}_{\varepsilon_1;\text{int}}^{[1];M} & \mathbf{E}_{\varepsilon_1;\text{int}}^{[1];N}
       \end{bmatrix}^\dagger,
   \label{eq:FDEprop_grad}
\end{eqnarray}
with $M \neq N$, what leads to a system of LR equations
\begin{align}
      (\mathbf{E}_{0}^{[2];M,M} + \mathbf{E}_{0;\text{int}}^{[2];M,M})\mathbf{X}_{\varepsilon_1}^{M}
      + \mathbf{E}_{0;\text{int}}^{[2];M,N}\mathbf{X}_{\varepsilon_1}^{N}
      &= -(\mathbf{E}_{\varepsilon_1}^{[1];M} + \mathbf{E}_{\varepsilon_1;\text{int}}^{[1];M}), \label{eq:lr_fde_I} \\
      \mathbf{E}_{0;\text{int}}^{[2];N,M}\mathbf{X}_{\varepsilon_1}^{M}
      + (\mathbf{E}_{0}^{[2];N,N} + \mathbf{E}_{0;\text{int} }^{[2];N,N})\mathbf{X}_{\varepsilon_1}^{N}
      &= -(\mathbf{E}_{\varepsilon_1}^{[1];N} + \mathbf{E}_{\varepsilon_1;\text{int}}^{[1];N}), \label{eq:lr_fde_II}
\end{align}
where the response vector has also been split into blocks pertaining to each
subsystem, $\mathbf{X}_{\varepsilon_1} = [\mathbf{X}_{\varepsilon_1}^{M} \quad \mathbf{X}_{\varepsilon_1}^{N}]^\dagger$.
The matrix elements of each sub-block have the form
\begin{equation}
\mathbf{E}_{0}^{[2];M,M} = \frac{\partial^2E_{M}}{\partial \kappa^{M}_{pq} \partial \kappa^{M}_{rs}};\qquad \qquad
\mathbf{E}_{0;\text{int}}^{[2];M,N} = \frac{\partial^2E_\text{int}}{\partial \kappa^{M}_{pq} \partial \kappa^{N}_{rs}}
\end{equation}
for the Hessian and
\begin{equation}
\mathbf{E}_{\varepsilon_1}^{[1];M} = \frac{\partial^2 E_{M}}{\partial \kappa^{M}_{ai} \partial \varepsilon_1};\qquad \qquad
\mathbf{E}_{\varepsilon_1;\text{int}}^{[1];M} = \frac{\partial^2 E_\text{int}}{\partial \kappa^{M}_{ai} \partial \varepsilon_1}
\end{equation}
for the property gradient.

While the subsystem contributions to the electronic Hessian and property gradient are as in (S)DFT,
the interaction contributions are calculated from the chain rule, employing the parametrization of Eq.~\ref{eq:rhotot_parametrization},
which allows for a straightforward separation of the contributions from perturbed densities of both subsystems ($M, N \in \{I, II\}$):
 \begin{align}
    \frac{\partial^2 E_\text{int}}{\partial \kappa_{pq}^{M} \partial \kappa_{rs}^N}\biggr|_0
    &=  \delta_{MN}
        \int
            \frac{\delta E_\text{int}}{\delta \rho_k^M}
            \frac{\partial^2 \rho_k^M}{\partial \kappa_{pq}^M \partial \kappa_{rs}^N}\biggr|_0
    +
        \iint
           \frac{\delta^2 E_\text{int}}{\delta \rho_k^M \delta \rho_{k'}^N}
           \frac{\partial \rho_k^M}{\partial \kappa_{pq}^M}\biggr|_0
           \frac{\partial \rho_{k'}^N}{\partial \kappa_{rs}^N} \biggr|_0
    \label{eq:d2_Eint_dkappa_dkappa}   \\
    \frac{\partial^2 E_\text{int}}{\partial \kappa_{pq}^{M} \partial \varepsilon_1}\biggr|_0
    &=  \int
           \frac{\delta E_\text{int}}{\delta \rho_k^M}
           \frac{\partial^2 \rho_k^M}{\partial \kappa_{pq}^M \partial \varepsilon_1}\biggr|_0
    +
         \iint
            \frac{\delta^2 E_\text{int}}{\delta \rho_k^M \delta \rho_{k'}^N}
            \frac{\partial \rho_k^M}{\partial \kappa_{pq}^M}\biggr|_0
            \frac{\partial \rho_{k'}^N}{\partial \varepsilon_1} \biggr|_0
    \label{eq:d2_Eint_dkappa_dex}
\end{align}
The functional derivatives of the interaction energy with respect to the
general density components are the embedding potential $v_\text{emb;k}^M$ of
Eq.~\ref{eq:embpot} and the embedding kernel ($M, N \in \{I, II\}$):
    \begin{align}
    w^{M,N}_\text{emb;k,k'}(\vec{r}_1,\vec{r}_2) &=
    \frac{\delta^2 E_\text{int}}{\delta\rho_k^M (\vec{r}_1) \delta\rho_{k'}^N(\vec{r}_2)}
    = (1 - \delta_{MN}) \frac{1}{|\vec{r}_1 - \vec{r}_2|} \label{eq:embkernel} \\
      &\quad + \frac{\delta^2 E_\text{xck}}{\delta \rho_k^\text{tot}(\vec{r}_1) \delta \rho_{k'}^\text{tot}(\vec{r}_2)}
      - \delta_{MN} \frac{\delta^2 E_\text{xck}}{\delta \rho_k^M(\vec{r}_1) \delta \rho_{k'}^M(\vec{r}_2)}
      \nonumber
    \end{align}
We recall that in both cases the functional derivatives are calculated with the ground-state densities.

\paragraph*{Electronic Hessian}
The interaction energy contributions to the electronic Hessian
(Eq.\ref{eq:d2_Eint_dkappa_dkappa}) can be further rewritten as ($M \neq N$):
 \begin{align}
    \frac{\partial^2 E_\text{int}}{\partial \kappa_{pq}^{M} \partial \kappa_{rs}^N}\biggr|_0
    &=  \delta_{MN}
        \int
            v_\text{emb;k}^M
            \frac{\partial^2 \rho_k^M}{\partial \kappa_{pq}^M \partial \kappa_{rs}^N}\biggr|_0
    +
        \iint
           w^{M,M}_\text{emb;k,k'}
           \frac{\partial \rho_k^M}{\partial \kappa_{pq}^M}\biggr|_0
           \frac{\partial \rho_{k'}^M}{\partial \kappa_{rs}^M} \biggr|_0
    \label{eq:d2_Eint_dkappa_dkappa_v_w11} \\
    &+
        \iint
           w^{M,N}_\text{emb;k,k'}
           \frac{\partial \rho_k^M}{\partial \kappa_{pq}^M}\biggr|_0
           \frac{\partial \rho_{k'}^N}{\partial \kappa_{rs}^N} \biggr|_0
    \label{eq:d2_Eint_dkappa_dkappa_w12},
  \end{align}
discerning the embedding potential as well as the uncoupled and coupled
embedding kernel terms. In the current DIRAC
implementation\cite{hofener-jcp-136-044104-2012} only the terms from
Eq.~\ref{eq:d2_Eint_dkappa_dkappa_v_w11} are included, so that coupling
contributions of Eq.~\ref{eq:d2_Eint_dkappa_dkappa_w12} are neglected. As it is
usually the case, the Hessian is not explicitly constructed but rather its
eigenvectors and eigenvalues are obtained by iterative
approaches.\cite{saue-jcp-118-522-2003}

\paragraph*{Property gradient}
When there is no dependence of $\rho_k$ on the perturbation,
Eq.~\ref{eq:d2_Eint_dkappa_dex} is identically zero and the property gradient
contains only contributions from the isolated subsystems. As terms of
Eq.~\ref{eq:d2_Eint_dkappa_dex} are non-zero when LAOs are used, from now on
they will be referred to as FDE-LAO contributions to the property gradient.
Eq.~\ref{eq:d2_Eint_dkappa_dex} can be rewritten in a more explicit form ($M
\neq N$),
  \begin{align}
    \frac{\partial}{\partial B_{\alpha}}
    \frac{\partial E_\text{int}}{\partial \kappa_{ai}^M}\biggr|_0
   &=
    -\int
    v_\text{emb;k}^{M}(\vec{r})
    \breve{\Omega}_{ia;k}^{B_{\alpha};M}
    \text{d} \vec{r}
    \label{eq:fde_prop_grad1} \\
    &- \iint
    w_\text{emb;k,k'}^{M, M}(\vec{r}_1, \vec{r}_2)
    \Omega_{ia;k}^M (\vec{r}_1)
    \breve{\Omega}_{jj;k'}^{B_{\alpha};M}(\vec{r}_2)
    \text{d} \vec{r}_1
    \text{d} \vec{r}_2
    \label{eq:fde_prop_grad2}\\
    &- \iint
    w_\text{emb;k,k'}^{M, N}(\vec{r}_1, \vec{r}_2)
    \Omega_{ia;k}^M (\vec{r}_1)
    \breve{\Omega}_{jj;k'}^{B_{\alpha};N}(\vec{r}_2)
    \text{d} \vec{r}_1
    \text{d} \vec{r}_2.
    \label{eq:fde_prop_grad3}
  \end{align}
employing the notation for the embedding potential (Eq.~\ref{eq:embpot}), the
embedding kernel (Eq.~\ref{eq:embkernel}) and the derivatives of orbital
overlap distributions (summarized in Table~\ref{tab:density_derivs}). Detailed
working expressions used for practical implementation of
Eq.~\ref{eq:fde_prop_grad1}-\ref{eq:fde_prop_grad3} are presented in Appendix
(Eq.~\ref{eq:fde_lao_prop_grad_appendix}).

Terms dependent on one subsystem,
Eqs.~\ref{eq:fde_prop_grad1}-\ref{eq:fde_prop_grad2}, are in effect analogous
to the XC contributions to the property gradient in OMO
basis,\cite{olejniczak-jcp-136-014108-2012} only with derivatives of the
interaction energy replacing derivatives of the XC energy.  In the LR
algorithm\cite{saue-jcp-118-522-2003} the property gradient is calculated once
and is not updated in the iterative procedure, therefore the computational cost
of including FDE-LAO terms does not significantly increase an overall cost of
calculations, unless the coupling terms (Eq.~\ref{eq:fde_prop_grad3}) are
considered, what will be briefly discussed in the following.

\paragraph*{Coupling kernel contributions to the linear response function} As
terms dependent on the embedding kernel (Eq.~\ref{eq:embkernel}) may involve
perturbed densities of two different subsystems ($M \neq N$), they will
introduce a coupling between these subsystems through Coulomb and non-additive
terms.

The coupling kernel contributions to the electronic Hessian
(Eq.~\ref{eq:d2_Eint_dkappa_dkappa_w12}) have been discussed at length in the
context of excitation
energies\cite{neugebauer-jcp-126-134116-2007,hofener-jcp-136-044104-2012,hofener-jctc-12-549-2016,pavanello-jcp-138-204118-2013,artiukhin-jcp-142-234101-2015}
or electric polarizabilities,\cite{neugebauer-jcp-131-084104-2009} and were
shown to be important for extensive properties or when excitations cannot be
considered (to good accuracy) as dominated by local components, but can often
be neglected otherwise.\cite{neugebauer-jpca-109-7805-2005,
jacob-jcp-125-194104-2006,neugebauer-jpca-110-8786-2006,bulo-jpca-112-2640-2008,gomes-pccp-10-5353-2008,gomes-pccp-15-15153-2013,neugebauer-jcp-122-094115-2005}

The coupling kernel contributions to the property gradient have received less
attention so far. As all FDE-LAO contributions to the property gradient result
from using LAOs, which shift the gauge origin from an arbitrary point to the
center of nuclei, this coupling term can be regarded as a small correction due
to the shift from the origin -- e.g. in the center of mass of subsystem $I$ --
to the centers of nuclei in subsystem $II$. While we still lack the coupling
contributions to the Hessian, in the following we shall nevertheless
investigate the relative importance of this term in the property gradient.

The presence of coupling terms increases the complexity and cost of
calculations.  As LR equations are solved for one subsystem at a time, the
necessary derivatives of the density of subsystem $II$ have to be calculated
and stored before the response equations for subsystem $I$ are invoked.
Eq.~\ref{eq:fde_prop_grad3} involves the non-additive $E_\text{xck}$ term
(calculated analogously as the uncoupled $E_\text{xck}$ part), as well as the
Coulomb contribution,
  \begin{equation}
    \int w_\text{emb;0}^{I,II;\text{Coulomb}}
    \breve{\Omega}_{jj;0}^{B_{\alpha};II}(\vec{r}_1)
    \Omega_{ia;0}^I(\vec{r})
    \text{d} \vec{r}_1,
  \end{equation}
which in our implementation is calculated via numerical integration of the expression
\begin{equation}
\int \left[\sum_{\mu\nu} c_{\mu i}^{\star} c_{\nu a} \int \frac{\chi_\mu^{I,\dagger}(\vec{r}_1) \Sigma_0 \chi_\nu^I (\vec{r}_1)}{|\vec{r} - \vec{r}_1|}d\vec{r}_1 \right]\left[\breve{\Omega}_{jj,0}^{B_{\alpha}, II} (\vec{r}s)\right]  d\vec{r} .
\end{equation}

\paragraph{Expectation values}

In a manner similar to the total energy in Eq.~\ref{eq:Etot}, the expectation
value term (second term in Eq.~\ref{eq:d2E_dYdY}) can be subdivided into
subsystem and interaction parts:
  \begin{equation}
  \frac{\partial^2 E}{\partial \varepsilon_1 \partial \varepsilon_2} \biggr|_0
    = \frac{\partial^2E_I}{\partial \varepsilon_1 \partial \varepsilon_2} \biggr|_0 + \frac{\partial ^2E_{II}}{\partial \varepsilon_1 \partial \varepsilon_2} \biggr|_0 + \frac{\partial ^2E_\text{int}}{\partial \varepsilon_1 \partial \varepsilon_2} \biggr|_0. \label{eq:general_subsystem_expval}
  \end{equation}
The subsystem contributions will have the same form as discussed elsewhere, whereas the interaction term is
 \begin{align}
    \frac{\partial^2 E_\text{int}}{\partial \varepsilon_1 \partial \varepsilon_2}\biggr|_0
    &= \sum_{M= I, II}
        \int
            \frac{\delta E_\text{int}}{\delta \rho_k^M}
            \frac{\partial^2 \rho_k^M}{\partial \varepsilon_1 \partial \varepsilon_2}\biggr|_0
            \label{eq:d2_Eint_depsilonA_depsilonB_1}  \\
    &+
        \sum_{M, N = I, II} \iint
           \frac{\delta^2 E_\text{int}}{\delta \rho_k^M \delta \rho_{k'}^N}
           \frac{\partial \rho_k^M}{\partial \varepsilon_1}\biggr|_0
           \frac{\partial \rho_{k'}^N}{\partial \varepsilon_2}\biggr|_0 .
    \label{eq:d2_Eint_depsilonA_depsilonB_2}
\end{align}
From this expression one can, as in the linear response case, distinguish interaction contributions to each of
subsystems,
  \begin{align}
    \frac{\partial^2 E^{M}_\text{int}}
         {\partial \varepsilon_1 \partial \varepsilon_2}\biggr|_0
    &=  \int
            \frac{\delta E_\text{int}}{\delta \rho_k^M}
            \frac{\partial^2 \rho_k^M}{\partial \varepsilon_1 \partial \varepsilon_2}\biggr|_0
    +
        \iint
           \frac{\delta^2 E_\text{int}}{\delta \rho_k^M \delta \rho_{k'}^M}
           \frac{\partial \rho_k^M}{\partial \varepsilon_1}\biggr|_0
           \frac{\partial \rho_{k'}^M}{\partial \varepsilon_2} \biggr|_0
           \nonumber \\
    &=
       \int
           v_\text{emb;k}^{M}(\vec{r})
           \breve{\Omega}_{ii;k}^{B_{\alpha}B_{\beta};M}
           \text{d} \vec{r}
           \label{eq:d2_Eint_depsilonA_depsilonB_subsystem_i_v}\\
    &+
       \iint
           w_\text{emb;k,k'}^{M, M}(\vec{r}_1, \vec{r}_2)
           \breve{\Omega}_{ii;k}^{B_{\alpha};M} (\vec{r}_1)
           \breve{\Omega}_{jj;k'}^{B_{\beta};M}(\vec{r}_2)
           \text{d} \vec{r}_1
           \text{d} \vec{r}_2,
           \label{eq:d2_Eint_depsilonA_depsilonB_subsystem_i_w}
  \end{align}
containing embedding potential and kernel contributions, and those which depend on both subsystems ($M \neq N$)
\begin{align}
    \frac{\partial^2 E^{MN}_\text{int}}
         {\partial \varepsilon_1 \partial \varepsilon_2}\biggr|_0
    &=  \iint
           \frac{\delta^2 E_\text{int}}{\delta \rho_k^M \delta \rho_{k'}^N}
           \frac{\partial \rho_k^M}{\partial \varepsilon_1}\biggr|_0
           \frac{\partial \rho_{k'}^N}{\partial \varepsilon_2} \biggr|_0
           \nonumber   \\
     &=
        \iint
        w_\text{emb;k,k'}^{M,N}(\vec{r}_1, \vec{r}_2)
        \breve{\Omega}_{ii;k}^{B_{\alpha};M} (\vec{r}_1)
        \breve{\Omega}_{jj;k'}^{B_{\beta};N}(\vec{r}_2)
        \text{d} \vec{r}_1
        \text{d} \vec{r}_2,
        \label{eq:d2_Eint_depsilonA_depsilonB_subsystems_ij_w}
\end{align}
made-up exclusively of a coupling kernel term. Working equations are presented
in the Appendix (Eq.~\ref{eq:fde_lao_expval_appendix}).

\paragraph*{Coupling kernel contributions to the expectation value}
As in the case of property gradient, all terms in
Eqs.~\ref{eq:d2_Eint_depsilonA_depsilonB_1} -
\ref{eq:d2_Eint_depsilonA_depsilonB_2} would be zero in
perturbation-independent basis sets, thus from now on they will be referred to
as FDE-LAO contributions to the expectation value.  Although
Eq.~\ref{eq:d2_Eint_depsilonA_depsilonB_subsystems_ij_w} depends on the
coupling kernel, it does not involve the relaxation of the subsystems'
densities as in the electronic Hessian but rather the static correction to the
choice of response parameters. Here we note that first-quantization and
second-quantization formulations of second-order magnetic properties in LAO
basis define expectation value contributions
differently.\cite{ilias-jcp-131-124119-2009}

\subsection{Tensor Expressions for the molecular properties and their representation in terms of magnetically induced currents}
\label{subsec:tensor}

The theory discussed above is sufficient to determine the properties of
interest in the subsystem approach. However, these properties can also be
presented in a different mathematical form using the linearity of the 4c DC
Hamiltonian in applied perturbations, complemented by the formulation involving
magnetically induced current densities, which more directly conveys the
physical characteristics of each property.

\subsubsection{NMR shielding and indirect spin-spin coupling tensors}
\label{subsub:tensor_shield}
The NMR shielding or the NMR indirect spin-spin coupling tensors in
Eqs.~\ref{eq:spinspin_simple} and~\ref{eq:shield_simple} can be recast in a
computationally advantageous form\cite{ilias-jcp-131-124119-2009} in terms of
expectation values involving the hyperfine operator for a nuclei $L$, the
unperturbed spinors $| \psi_i \rangle$ and the first-order perturbed
spinors\cite{ilias-jcp-131-124119-2009,olejniczak-jcp-136-014108-2012} $|
\tilde{\psi}_i^{\epsilon_i} \rangle$, yielding the general expression
  \begin{equation}
    M^{\epsilon;L}_{\alpha\beta} = \sum_i \left\{ \langle \tilde{\psi}_i^{\epsilon_\alpha} | \hat{h}_{m_{L;\beta}} | \psi_i \rangle + \langle \psi_i | \hat{h}_{m_{L;\beta}} | \tilde{\psi}_i^{\epsilon_\alpha} \rangle \right\}.
    \label{eq:nmr_tensors_generall_working}
  \end{equation}
The expression for the shielding tensor $\sigma^L_{\alpha\beta}$ is therefore
obtained from Eq.~\ref{eq:nmr_tensors_generall_working} by employing the spinors
perturbed by the external magnetic field ($\epsilon = \vec{B}$), $|
\tilde{\psi}_i^{B_\alpha} \rangle$, and by the same token the spin-spin
coupling tensor $K^{KL}_{\alpha\beta}$ is obtained by employing the spinors
perturbed by the nuclear magnetic dipole ($\epsilon = \vec{m}_K$),  $|
\psi_i^{m_{K;\alpha}}\rangle$.

In the FDE case, as each subsystem is described by its own set of
externally-orthogonal orbitals, we can rewrite the expression in
Eq.~\ref{eq:nmr_tensors_generall_working} as
\begin{eqnarray}
 M^{\epsilon;L}_{\alpha\beta}
   &=& \sum_{i\in I}
        \left\{
         \langle \tilde{\psi}_i^{\epsilon_\alpha} | \hat{h}_{m_{L;\beta}} | \psi_i \rangle + \langle \psi_i | \hat{h}_{m_{L;\beta}} | \tilde{\psi}_i^{\epsilon_\alpha} \rangle
        \right\}
        \label{eq:nmr_tensors_general_working_subsystem_I} \\
   &+& \sum_{j\in II}
        \left\{
         \langle \tilde{\psi}_j^{\epsilon_\alpha} | \hat{h}_{m_{L;\beta}} | \psi_j \rangle + \langle \psi_j | \hat{h}_{m_{L;\beta}} | \tilde{\psi}_j^{\epsilon_\alpha} \rangle
        \right\}.
        \label{eq:nmr_tensors_general_working_subsystem_II}
\end{eqnarray}
The FDE expression for $\sigma^L_{\alpha\beta}$ or $K^{KL}_{\alpha\beta}$ can
be further approximated by neglecting the terms arising from
Eq.~\ref{eq:nmr_tensors_general_working_subsystem_II}. In the case of NMR
shieldings, assuming nucleus $L$ belongs to subsystem $I$, this approximation
should be sufficient, especially if the overlap between two subsystems is small,
but whatever the case we can estimate this missing contribution by the
magnetically-induced current density formulation outlined in
section~\ref{subsubsec:currents}. For the spin-spin tensor this approximation
should also be good due to the local nature of the hyperfine operator, if both
$K$ and $L$ belong to subsystem $I$ (a restriction in our current
implementation).

\subsubsection{Magnetizability tensor}
\label{subsub:magn}
Contrary to NMR properties, the magnetizability tensor is not a local property
as the Zeeman operator (Eq.~\ref{eq:Zeeman_hyperfine}) affects the whole
system. It can be expressed in terms of the sum of (interacting)
intra-subsystem and inter-subsystem contributions
\begin{equation}
\xi_{\alpha\beta} = \xi_{\alpha\beta}^{I,(II)} + \xi_{\alpha\beta}^{II,(I)},
\end{equation}
where
 \begin{align}
    \xi_{\alpha\beta}^{I,(II)}
    & = \left[
           \frac{\partial^2E_I}{\partial B_\alpha \partial B_\beta} \biggr|_0
         + \frac{\partial ^2E_\text{int}^{M = I}}{\partial B_\alpha \partial B_\beta} \biggr|_0
      \right]
      \label{eq:magn_expval_I} \\
    &+
         \left[
           \frac{\partial^2 E_I}{\partial \kappa_{pq}^I \partial B_\beta}
           +
           \frac{\partial^2 E_\text{int}}{\partial \kappa_{pq}^I \partial B_\beta}
         \right]\frac{\partial \kappa_{pq}^I}{\partial B_\alpha}.
     \label{eq:magn_rsp_I}
  \end{align}
The terms in Eq.~\ref{eq:magn_rsp_I} are calculated by solving the linear
response equations for subsystems $I$ with FDE-LAO contributions to the
property gradient (Eqs.~\ref{eq:fde_prop_grad1} - \ref{eq:fde_prop_grad3}).
The term involving $E_\text{int}^{M = I}$ in Eq.~\ref{eq:magn_expval_I} will contain
intra-subsystem contributions of
Eqs.~\ref{eq:d2_Eint_depsilonA_depsilonB_subsystem_i_v} and
\ref{eq:d2_Eint_depsilonA_depsilonB_subsystem_i_w} with the summation
restricted to subsystem $I$ and the inter-subsystem contribution from
Eq.~\ref{eq:d2_Eint_depsilonA_depsilonB_subsystems_ij_w}, where the summation
is constrained to have $M = I$.  The second term in Eq.~\ref{eq:magn_rsp_I},
$\xi_{\alpha\beta}^{II,(I)}$, is obtained by permutation of indices $I$ and
$II$.

\subsubsection{Tensors in terms of induced currents}
\label{subsubsec:currents}

The relativistic current density vector and its first-order derivatives with respect to perturbations,\cite{bast-cp-356-187-2009}
  \begin{equation}
    \vec{j} (\vec{r}) = -e \sum_i \psi_i^\dagger  c \vec{\alpha} \psi_i
    \label{eq:j}
  \end{equation}
  \begin{equation}
    \vec{j}^{\,\varepsilon_1} (\vec{r}) = -e \sum_i \left\{ \tilde{\psi}_i^{\varepsilon_1; \dagger}  c \vec{\alpha} \psi_i  + \psi_i^\dagger  c \vec{\alpha} \tilde{\psi}_i^{\varepsilon_1}  \right\},
    \label{eq:j^B}
  \end{equation}
allow to construct property densities,\cite{jameson-jcp-73-5684-1980} which may
be visualized on a grid and integrated -- giving the value of the corresponding
property.  Thus, the properties studied in this paper can be written as:
  \begin{equation}
    \sigma^K_{\alpha\beta} = -\frac{1}{c^2} \int \frac{1}{r_K^3} \left( \vec{r}_K \times \vec{j}^{\,B_\alpha} \right)_\beta d\vec{r},
    \label{eq:shielding_current}
  \end{equation}
  \begin{equation}
    K^{KL}_{\alpha\beta} = - \frac{1}{c^2} \int \frac{1}{r_K^3} \left( \vec{r}_K \times \vec{j}^{\,m_{L;\alpha}} \right)_\beta d\vec{r}.
    \label{eq:spinspin_current}
  \end{equation}
    \begin{equation}
    \xi_{\alpha\beta} = - \frac{1}{2} \int \left( \vec{r}_G \times \vec{j}^{\,B_{\alpha}} \right)_\beta d\vec{r}.
    \label{eq:magn_current}
  \end{equation}
where the first-order current density perturbed by an external magnetic field
is calculated with LAOs.\cite{sulzer-pccp-13-20682-2011}

An advantage of using the induced current density is that while evaluated for
one subsystem, it can be contracted with the position vector pointing to the
other subsystem allowing to evaluate contributions, for example from
Eq.~\ref{eq:nmr_tensors_general_working_subsystem_II}, for the NMR shielding
tensor as: \begin{equation} \sigma^{K;II}_{\alpha\beta} = -\frac{1}{c^2} \int
\frac{1}{(\vec{r}_i^{II} - \vec{R}_K^I)^3} \left( (\vec{r}_i^{II} -
\vec{R}_K^I) \times \vec{j}^{\,B_\alpha;II} \right)_\beta d\vec{r},
\label{eq:shielding_current_II} \end{equation} where the superscripts I and II
denoting the subsystems are written explicitly for each vector.
Eq.~\ref{eq:shielding_current_II} is analogous to nucleus independent chemical
shift (NICS) calculations outlined for FDE by Jacob and
Visscher.\cite{jacob-jcp-125-194104-2006}

\section{Computational details}
\label{sec:compdet}

We have investigated three hydrogen-bonded HXH--OH$_2$ complexes, where X = Se,
Te, Po. Their structures were optimized in ADF software,\cite{ADF14} using the
scalar version of the zeroth-order regular approximation
(ZORA)\cite{vanlenthe-jcp-99-4597-1993, vanlenthe-jcp-101-9783-1994}
Hamiltonian, the B3LYP\cite{becke-jcp-98-5648-1993} functional and basis sets
of the triple-zeta quality (TZ2P).\cite{vanlenthe-jcc-24-1142-2003} The
optimized structures are included as supplementary material. The structures for
the subsystems are taken from supermolecules without any further
reoptimization, so that calculation on isolated fragments can be though of as
equivalent to QM/MM embedding where only mechanical (``ME'') coupling between the
subsystems is taken into account.\cite{gomes-arc-108-222-2012}

The wave function optimization and magnetic properties calculations performed
with a development version of the DIRAC code\cite{DIRAC15} employed the DC
Hamiltonian and the PBE\cite{perdew-prl-77-3865-1996,perdew-prl-78-1396-1997}
functional. In the FDE calculations the non-additive exchange-correlation and
kinetic energy contributions were calculated with the PBE and
PW91k\cite{lembarki-pra-50-5328-1994} functionals, respectively. In
response calculations we have used the full derivatives of the PBE
and PW91k functionals provided by the XCFun
library.\cite{ekstrom-jctc-6-1971-2010} The basis sets were of
augmented triple-zeta quality: aug-cc-pVTZ\cite{dunning-jcp-90-1007-1989} for H
and O and dyall.acv3z\cite{dyall-tca-108-335-2002,dyall-tca-115-441-2006} for
X.

Calculations of NMR properties with the spin-orbit ZORA Hamiltonian (ZORA-SO)
were performed in ADF with the TZ2P basis set and the PBE functional. For FDE
calculations with ADF we have also employed the PBE and PW91k functionals for
the non-additive contributions.

In both DIRAC and ADF the Gaussian model of nuclear charge
distribution\cite{visscher-adandt-67-207-1997} was used and in the case of
DIRAC the $(SS|SS)$ class of two-electron integrals was replaced by a standard
correction.\cite{visscher-tca-98-68-1997} Also, in both cases we performed two
sets of FDE calculations, one using densities obtained for the isolated
subsystems as frozen densities (hereafter referred to as ``FDE(0)'') and
another where we optimized both subsystem densities by exchanging their role as
frozen/active densities in the ``freeze-thaw'' procedure, which was stopped
after 4 iterations with both densities fully optimized (hereafter referred to
as ``FDE(4)''). In tables in this paper we present only the latter results,
while full tables are available in ESI.

We note that the choice of PBE was motivated by minimizing the differences in
the computational setup between supermolecular and FDE calculations, so the
only additional approximation in the FDE case comes from the kinetic energy
functional. Since our aim is to compare supermolecular and FDE results, a
thorough study of the performance of different functionals
(exchange-correlation and/or kinetic) is beyond the scope of this paper.

We use the definitions of Mason\cite{mason-ssnmr-2-285-1993} for the isotropic
and the anisotropic parts of a tensor $\Omega$ in principal axis system, where
$\Omega_{33} \geq \Omega_{22} \geq \Omega_{11}$,
\begin{eqnarray}
\Omega_\text{iso}   &=& 1/3(\Omega_{11} + \Omega_{22} + \Omega_{33}) \\
\Omega_\text{aniso} &=& \Omega_{33} - 1/2(\Omega_{11} + \Omega_{22}).
\end{eqnarray}

We have also calculated and plotted NMR shielding densities and their differences.
In this paper we present the differential NMR shielding densities calculated by
subtracting the shielding density of nucleus X and H$_b$ in
Figure~\ref{fig:shielddens} arising from the induced current in both subsystems
(employing NICS method for the frozen subsystem) from the corresponding
shielding density in a supermolecule.  The shielding density is plotted on a
80$\times$80$\times$80 grid of points generated for supermolecules. The plots
were made with Mayavi - the library for interactive scientific data
visualization and 3D plotting in Python.\cite{ramachandran2011mayavi} More
plots and relevant data can be found in ESI.

\section{Results and discussion}
\label{sec:results}
In what follows we shall present all embedding results relative to the
supermolecular ones.  Thus, for a given molecular property $P$ we define
absolute ($\Delta P^{f}$) and relative ($\delta P^{f} $) property shifts in a
general manner as
\begin{eqnarray}
   \Delta P^{f} &=& P^\text{super} - (P^{I;f} + P^{II;f}) \\
   \label{eq:prop_shifts_general_absolute}
   \delta P^{f} &=& \Delta P^{f} / P^\text{super},
   \label{eq:prop_shifts_general_relative}
  \end{eqnarray}
where $P^\text{super}$ corresponds to the supermolecular value of property $P$ and
$P^{i;f}$ denotes the contribution to the property from the subsystem $i$. The
latter is obtained for the isolated subsystem ($f = \text{ME}$) or using FDE ($f =
\text{FDE}(n)$, with $n=0, 4$). In the case of NMR shieldings and indirect spin-spin
couplings, which are essentially local to one of the subsystems, $\Delta P^{f}$
is well approximated by neglecting $P^{II;f}$ in
Eq.~\ref{eq:prop_shifts_general_absolute}.  In the case of FDE calculations, we
introduce additional notation in order to discern FDE-LAO contributions to the
property gradient (NMR shielding and magnetizability) and to the expectation
value (magnetizability), which can be either neglected ($\text{FDE}(n)[0]$), limited
to the embedding potential ($\text{FDE}(n)[v]$) or to the embedding potential and the
uncoupled kernel ($\text{FDE}(n)[v+w_{u}]$) or also incorporating the coupling kernel
($w_c$) terms ($\text{FDE}(n)[v + w_\text{all}]$).

\subsection{NMR shielding tensor}
\label{subsec:results_shield}

The DC calculations of isotropic and anisotropic parts of NMR shielding tensor
are summarized in Table~\ref{tab:shieldings-dirac}. We present only the results
for nuclei of the active subsystem (H$_2$X) and we distinguish between the
hydrogen involved in the hydrogen bond (H$_b$) and the other pointing away from
the water molecule.

\begin{table}[hbt]
  \setlength\belowcaptionskip{5pt}
  \scriptsize
  \caption{
           \label{tab:shieldings-dirac}
           Absolute DC isotropic and anisotropic shielding values ($\sigma_\text{iso}^\text{super}$ and $\sigma_\text{aniso}^\text{super}$, in ppm) of nuclei in H$_2$X (X = Se, Te, Po) subsystems in H$_2$X--H$_2$O, and absolute shifts ($\Delta \sigma$, in ppm) for the isolated (``ME'') and embedded (``FDE(4)'') H$_2$X molecules in the presence of H$_2$O. For FDE the
values for different approximations in the FDE-LAO treatment ($a: [0]$, $b: [v]$, $c: [v+w_u]$, $d: [v+w_\text{all}]$) are shown. 
          }
% \begin{tabular}{l rrrrrr r rrrrrr}
 \begin{tabular}{l rrrrrr}
  \toprule
Atom & 
$\sigma_{iso}^\text{super}$ &
$\Delta \sigma_\text{iso[a]}^\text{FDE(4)}$ &
$\Delta \sigma_\text{iso[b]}^\text{FDE(4)}$ &
$\Delta \sigma_\text{iso[c]}^\text{FDE(4)}$ &
$\Delta \sigma_\text{iso[d]}^\text{FDE(4)}$ &
$\Delta \sigma_\text{iso}^\text{ME}$ \\
&&&&&&\\
\hline
&&&&&&\\
Se    &  2378.03 &--100.75 & --12.54  & --12.55  & --12.33  &   38.25  \\ % 609.27  &   21.80  &  --0.13  &  --0.43  &  --0.19  &     0.17  \\
H$_b$ &    30.88 &  --0.39 &  --0.83  &  --0.67  &  --0.62  &  --2.09  \\ %  22.19  &    5.05  &    6.00  &    5.76  &    5.70  &     7.29  \\
H     &    33.42 &    0.24 &    0.02  &    0.03  &    0.03  &    0.43  \\ %  15.13  &  --1.17  &  --0.22  &  --0.22  &  --0.21  &     0.09  \\
      &          &         &          &          &          &          \\ %         &          &          &          &          &           \\
Te    &  4667.85 &--142.39 &  --4.62  &  --9.16  &  --8.88  &   67.48  \\ %1189.67  &   29.78  &    0.56  &  --1.22  &  --1.21  &     3.60  \\
H$_b$ &    35.62 &  --0.29 &  --0.68  &  --0.44  &  --0.42  &  --1.70  \\ %  14.59  &    0.50  &    0.02  &    0.24  &    0.28  &   --1.31  \\
H     &    37.85 &  --0.02 &    0.01  &    0.02  &    0.02  &    0.54  \\ %  15.24  &    0.54  &    0.09  &    0.08  &    0.08  &   --0.64  \\
      &          &         &          &          &          &          \\ %         &          &          &          &          &           \\
Po    & 13985.80 &--224.54 &   18.28  &  --3.52  &  --3.13  &  137.84  \\ %5556.67  &   61.88  & --15.12  & --18.36  & --18.61  & --463.51  \\
H$_b$ &    40.80 &    0.50 &    0.07  &  --0.09  &  --0.09  &  --0.57  \\ % 105.25  &    0.98  &  --1.28  &  --1.67  &  --1.67  &   --6.01  \\
H     &    42.29 &  --0.22 &  --0.03  &    0.00  &    0.00  &    0.89  \\ % 107.80  &    1.93  &    0.16  &    0.08  &    0.09  &   --4.03  \\
&&&&&&\\
\cline{2-7}
&&&&&&\\
     & 
$\sigma_\text{aniso}^\text{super}$ &
$\Delta \sigma_\text{aniso[a]}^\text{FDE(4)}$ &
$\Delta \sigma_\text{aniso[b]}^\text{FDE(4)}$ &
$\Delta \sigma_\text{aniso[c]}^\text{FDE(4)}$ &
$\Delta \sigma_\text{aniso[d]}^\text{FDE(4)}$ &
$\Delta \sigma_\text{aniso}^\text{ME}$ \\
&&&&&&\\
\cline{2-7}
&&&&&&\\
Se    &  609.27  &   21.80  &  --0.13  &  --0.43  &  --0.19  &     0.17  \\
H$_b$ &   22.19  &    5.05  &    6.00  &    5.76  &    5.70  &     7.29  \\
H     &   15.13  &  --1.17  &  --0.22  &  --0.22  &  --0.21  &     0.09  \\
      &          &          &          &          &          &           \\
Te    & 1189.67  &   29.78  &    0.56  &  --1.22  &  --1.21  &     3.60  \\
H$_b$ &   14.59  &    0.50  &    0.02  &    0.24  &    0.28  &   --1.31  \\
H     &   15.24  &    0.54  &    0.09  &    0.08  &    0.08  &   --0.64  \\
      &          &          &          &          &          &           \\
Po    & 5556.67  &   61.88  & --15.12  & --18.36  & --18.61  & --463.51  \\
H$_b$ &  105.25  &    0.98  &  --1.28  &  --1.67  &  --1.67  &   --6.01  \\
H     &  107.80  &    1.93  &    0.16  &    0.08  &    0.09  &   --4.03  \\
%\bottomrule
\end{tabular}
\normalsize
\end{table}

\subsubsection{Isolated subsystems}

From the isolated (``ME'') calculation we observe that the hydrogen-bonded water
strongly affects the isotropic and anisotropic parts of the NMR shielding
tensors of nuclei of the active subsystems, leading to the shielding of the
heavy centers and the deshielding of H$_b$, in agreement with established
observations on hydrogen-bonded systems.\cite{iupac-hbond-2011}

The values of $\Delta \sigma_\text{iso}^\text{ME}$ are progressively larger with the
increase of atomic number of X for all nuclei of an active subsystem: for the X
nuclei they range from 38 ppm for SeH$_2$ to 138 ppm for PoH$_2$, for the non
H-bonded hydrogen nuclei -- from 0.4 ppm for SeH$_2$ to 0.9 ppm for PoH$_2$ and
for H$_b$ nuclei -- from --2 ppm for SeH$_2$ to --0.6 ppm for PoH$_2$, which 
for hydrogen nuclei are
significant since $^1$H NMR shielding is between 10 ppm and 30 ppm in most
applications.\cite{nmr-book-kaupp}

While these $\Delta \sigma$ values are relatively small in comparison to the
absolute shieldings, they can nevertheless be significant in NMR experiments --
for instance, both $^{77}$Se and $^{125}$Te nuclei are known to be very
sensitive to the environment (e.g. solvent, its concentration and
temperature\cite{luthra-sete-2010}) and even though they span wide chemical
shift ranges (6000 ppm for
$^{77}$Se\cite{kemp-ssnmr-34-224-2008,duddeck-annrepnmr-52-105-2004} and 7000
ppm\cite{luthra-sete-2010} for $^{125}$Te), shifts of around 30 ppm (Se) or 60
ppm (Te) are fingerprints of a specific
solvent.\cite{luthra-sete-2010,rusakov-mrc-53-485-2015,rusakov-jcc-36-1756-2015,hayashi-rscadv-4-44795-2014}

\begin{figure}[!htb]
\flushright
\hfill
\minipage{\linewidth}
 \includegraphics[width=1.00\linewidth]{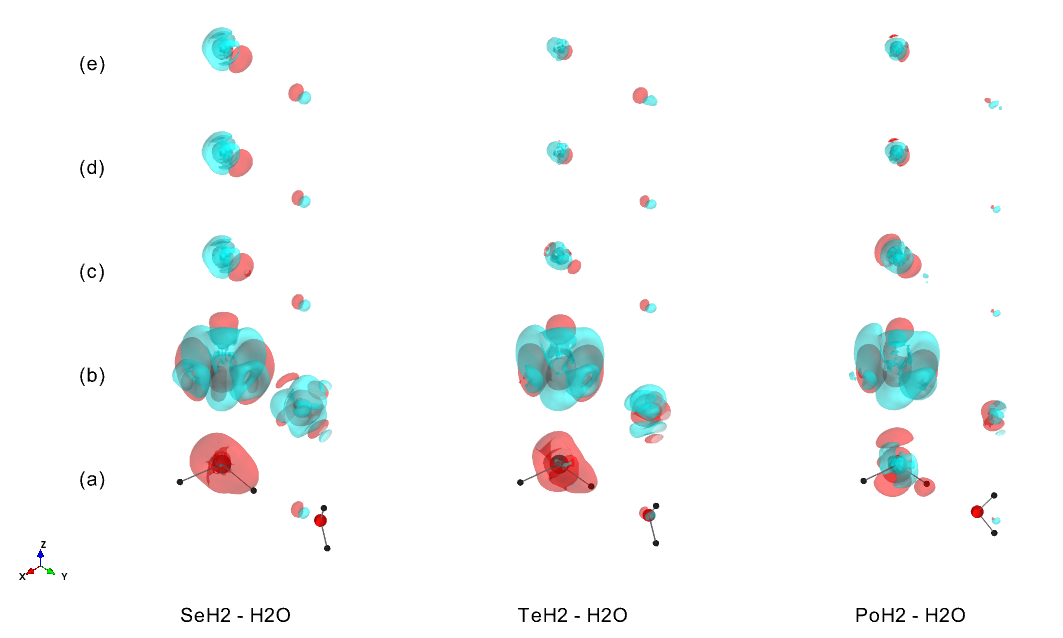}
%\captionof{figure}{X = Se, Te, Po differential isotropic shielding density isosurfaces at isovalues at +0.53 ppm (red) and -0.53 ppm\ (blue)}
\endminipage
\hfill
\minipage{\linewidth}
  \includegraphics[width=0.90\linewidth]{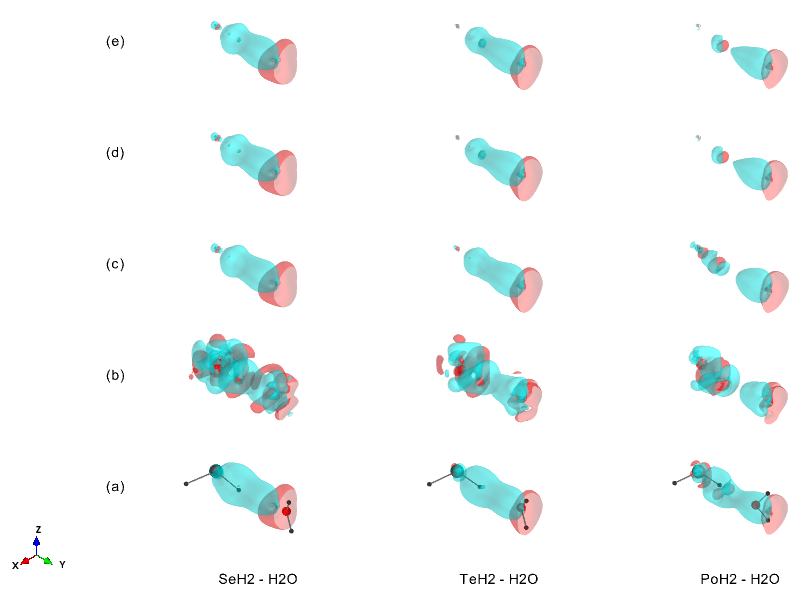}
%\captionof{figure}{H$_b$ differential isotropic shielding density isosurfaces at isovalues +0.53 ppm (red) and -0.53 ppm\ (blue)}
\endminipage
  \caption{Differential isotropic shielding density isosurfaces\cite{ourfigures} (isovalues at +0.53 ppm (red) and -0.53 ppm\ (blue)) for XH$_2$ - H$_2$O systems X = Se, Te, Po (top) and H$_b$ (bottom), calculated as a difference between supermolecule shielding densities and the sum of subsystem shielding densities approximated as: (a) ME (b) FDE(4)[0](c) FDE(4)[v] (d) FDE(4)[v+w$_u$] (e) FDE(4)[v+w$_\text{all}$]. Color of atoms: X (blue), O(red), H(grey).}
 \label{fig:shielddens}
\hfill
\end{figure}

For the anisotropies, isolated (``ME'') results are usually very different than the
supermolecular ones for H$_b$, with  $\Delta \sigma_\text{aniso}^\text{ME}$ of 7 ppm
for SeH$_2$, --1 ppm for TeH$_2$ and --6 ppm for PoH$_2$, which represent
deviations of 33\%, 9\% and 6\% respectively. This is not unexpected, since in
order to properly capture the directionality of the hydrogen bond, electronic
effects must be taken into account. Interestingly, these discrepancies are also
seen for the Po nucleus in PoH$_2$, whose shielding anisotropy differs from the
supermolecular value by 464 ppm (or 8\% difference to the supermolecular
value), whereas no significant deviations are seen for Se or Te.

These tendencies can be better seen in the plots of the differential isotropic
shielding density in Figure~\ref{fig:shielddens} (a) for the heavy centers and
H$_b$. Those figures exhibit positive (pink) and negative (blue) isosurfaces,
which, respectively, depict more shielded and more deshielded areas in a
supermolecule than in the embedded subsystems, and which upon integration give
the corresponding values of $\Delta \sigma_\text{iso}^\text{ME}$.

We observe from Figure~\ref{fig:shielddens} (a) that the plots are rather similar
for Se and Te nuclei, exhibiting  small negative isosurfaces centered on a
heavy nucleus, surrounded by much larger positive isosurface elongated on
X--H$_b$ bond. In case of Po the differential shielding density is represented
by much more complex isosurfaces around heavy center, as the negative
isosurface centered on Po is larger than observed for Se and Te and surrounded
by many well-separated positive lobes. This indicates that even though the
property shift $\Delta \sigma_\text{iso}^\text{ME}$(Po) turns out to be relatively modest
compared to the value of the absolute shielding, it is a result of shielded and
deshielded areas cancelling out upon integration, reflecting the intricate
interplay between environmental and relativistic effects, which are quite
different for Te and Se.

Towards the heavier neighbouring center, hydrogen nuclei experience larger HALA
effects\cite{autschbach-ch4-2013} (reflected by increasing absolute values of
$\sigma_\text{iso}^\text{super}$(H$_b$)) contributing to the shielding of H$_b$ nuclei,
and competing with the deshielding effect caused by the interaction with water
molecule.  We also observe the difference in the non hydrogen bonded hydrogen
shielding between PoH$_2$ and the other species, which could also be a
consequence of the HALA effect.

\subsubsection{Frozen Density Embedding}
\label{subsub:results_shield_fde}

In what follows we discuss the relative importance of the FDE-LAO contributions
($v, w_u, w_c$), both in terms of the values of isotropic and anisotropic
$\Delta \sigma^\text{FDE(4)}$ indices as well as from the plots of the differential
isotropic shielding density in Figure~\ref{fig:shielddens} (b)-(e).  We also calculated
contributions to shielding of X and H$_b$ from the frozen subsystem using the
induced current density formulation from Eq.~\ref{eq:shielding_current_II}, but
since for all approximations these were found to be smaller than the assumed
accuracy (0.01 ppm), they are not shown separately.

Starting with the calculation where no FDE-LAO terms are added to the property
gradient, the results of isotropic shieldings are much worse than the ``ME''
values for all nuclei considered. The same conclusion is drawn for anisotropic
shieldings of Se and Te nuclei, but not for Po nucleus, for which
$\Delta \sigma_\text{aniso}^\text{FDE(4)}$ value remains smaller than the one incorporating only
the mechanical coupling.

The inclusion of FDE-LAO$[v]$ contributions yields significant improvement over
the isolated values. For the heavy elements, 
$\Delta\sigma_\text{iso}^\text{FDE(4)}$ increases from
the very negative values obtained without FDE-LAO terms to still significant
negative values for Se (--13 ppm), less negative for Te (--4 ppm) and positive
for Po (18 ppm). For H$_b$ the inclusion of the potential acts in the opposite
direction and we observe relatively small decrease of 
$\Delta\sigma_\text{iso}^\text{FDE(4)}$
values, whereas for the other hydrogen there is little change.

Uncoupled kernel contributions ($v+w_u$) further improve results as they
partially offset the $v$ contribution and reduce the $\Delta\sigma_\text{iso}^\text{FDE(4)}$ to
rather small values. The $w_u$ correction is much more significant for Po than
for Te, and for Se only little improvement is seen. The reason for this
difference among elements is that the kernel terms introduce contributions from
the response of the spin-density which becomes more significant as the elements
become heavier.  This is also the reason why $w_u$ contributions affect
the shielding of H$_b$ atoms (spin-orbit mechanism of HALA effect). 
The uncoupled kernel also accentuates the
trend seen for $\Delta\sigma_\text{aniso}^\text{FDE(4)}$(X), making 
it more negative and therefore
overestimating shielding anisotropy of X. This overestimation is also observed
for the $\sigma_\text{aniso}^\text{FDE(4)}$(H$_b$) except in SeH$_2$.

The coupling terms ($w_c$), on the other hand, act in general to offset the
uncoupled kernel terms but their magnitude is, as expected from the local
nature of the NMR shielding, much smaller than the latter for all atoms so it
plays no significant role in either the isotropic shielding value or the
shielding anisotropy.

\begin{table}[hbt]
  \setlength\belowcaptionskip{5pt}
  \scriptsize
  \caption{
           \label{tab:magnetizability}
           DC isotropic and first anisotropic magnetizabilities ($\xi_\text{iso}$ and $\xi_\text{aniso1}$, in SI units) for the H$_2$X--H$_2$O (X = Se, Te, Po) systems 
           as well as for the H$_2$X and H$_2$O subsystems, the latter as  isolated (``ME'') and embedded (``FDE(4)'') calculations. In the case of subsystem calculations
           the total $\xi^\text{tot}_\text{iso}$ and $\xi^\text{tot}_\text{aniso1}$ are given as the sum of the subsystem values. For FDE the values for the different approximations in the FDE-LAO 
           treatment ($a: [0]$, $b: [v]$, $c: [v+w_u]$, $d: [v+w_\text{all}]$) are shown.
          }
 \begin{tabular}{l rrrrrr}
  \toprule
System &
$\xi_\text{iso}^\text{super}$ &
$\xi_\text{iso[a]}^\text{FDE(4)}$ &
$\xi_\text{iso[b]}^\text{FDE(4)}$ &
$\xi_\text{iso[c]}^\text{FDE(4)}$ &
$\xi_\text{iso[d]}^\text{FDE(4)}$ &
$\xi_\text{iso}^\text{ME}$ \\
&&&&&&\\
\hline
&&&&&&\\
    SeH$_2$   &   ---    & --183.07 & --608.99  & --606.54  & --606.54  &  --602.19  \\%   ---    & --358.80  &  --47.12  &   --45.88  &  --45.88   &  --45.97  \\
    H$_2$O    &   ---    &   781.33 & --233.79  & --233.78  & --233.76  &  --234.19  \\%   ---    &--1396.29  &   --5.76  &    --5.76  &   --5.78   &   --6.51  \\
$\xi^\text{tot}$&--836.26&   598.26 & --842.77  & --840.32  & --840.30  &  --836.31  \\% --57.94  &--1755.08  &  --52.88  &   --51.64  &  --51.66   &  --52.48  \\
$\Delta \xi$  &    0.0   &--1434.52 &     6.51  &     4.06  &     4.04  &      0.05  \\%    0.0   &  1697.14  &   --5.07  &    --6.31  &   --6.28   &   --5.47  \\
              &          &          &           &           &           &            \\
    TeH$_2$   &   ---    & --630.07 & --858.94  & --848.68  & --848.74  &  --842.57  \\%   ---    & --227.24  &  --86.32  &   --80.29  &  --80.32   &  --80.30  \\
    H$_2$O    &   ---    &   235.10 & --233.74  & --233.74  & --233.60  &  --233.83  \\%   ---    & --650.52  &   --5.53  &    --5.53  &   --5.70   &   --6.06  \\
$\xi^\text{tot}$&-1080.67& --394.97 &--1092.69  &--1082.42  &--1082.33  & --1076.39  \\% --81.63  & --877.77  &  --91.85  &   --85.82  &  --86.02   &  --86.36  \\
$\Delta \xi$  &    0.0   & --685.71 &    12.01  &     1.74  &     1.66  &    --4.28  \\%    0.0   &   796.14  &    10.23  &      4.19  &     4.39   &     4.73  \\
              &          &          &           &           &           &            \\
    PoH$_2$   &  ---     & --895.55 &--1030.19  & --949.80  & --949.71  &  --940.09  \\%  ---     & --290.25  & --261.88  &   --91.58  &  --91.56   &  --91.79  \\
    H$_2$O    &  ---     & --169.52 & --234.11  & --234.10  & --233.18  &  --234.02  \\%  ---     &  --93.62  &   --6.45  &    --6.45  &   --7.47   &   --6.73  \\
$\xi^\text{tot}$&-1184.04&--1065.08 &--1264.30  &--1183.91  &--1182.89  & --1174.11  \\% --89.81  & --383.87  & --268.34  &   --98.03  &  --99.02   &  --98.52  \\
$\Delta \xi$  &    0.0   & --118.96 &    80.26  &   --0.13  &   --1.15  &    --9.92  \\%    0.0   &   294.06  &   178.53  &      8.22  &     9.21   &     8.71  \\
&&&&&&\\
\cline{2-7}
&&&&&&\\
&
$\xi_\text{aniso1}^\text{super}$ &
$\xi_\text{aniso1[a]}^\text{FDE(4)}$ &
$\xi_\text{aniso1[b]}^\text{FDE(4)}$ &
$\xi_\text{aniso1[c]}^\text{FDE(4)}$ &
$\xi_\text{aniso1[d]}^\text{FDE(4)}$ &
$\xi_\text{aniso1}^\text{ME}$ \\
&&&&&&\\
\cline{2-7}
&&&&&&\\
    SeH$_2$   &    ---    & --358.80  &  --47.12  &   --45.88  &  --45.88   &  --45.97  \\
    H$_2$O    &    ---    &--1396.29  &   --5.76  &    --5.76  &   --5.78   &   --6.51  \\
$\xi^\text{tot}$& --57.94 &--1755.08  &  --52.88  &   --51.64  &  --51.66   &  --52.48  \\
$\Delta \xi$  &     0.0   &  1697.14  &   --5.07  &    --6.31  &   --6.28   &   --5.47  \\
              &           &          &           &           &           &            \\
    TeH$_2$   &    ---    & --227.24  &  --86.32  &   --80.29  &  --80.32   &  --80.30  \\
    H$_2$O    &    ---    & --650.52  &   --5.53  &    --5.53  &   --5.70   &   --6.06  \\
$\xi^\text{tot}$& --81.63 & --877.77  &  --91.85  &   --85.82  &  --86.02   &  --86.36  \\
$\Delta \xi$  &     0.0   &   796.14  &    10.23  &      4.19  &     4.39   &     4.73  \\
              &           &          &           &           &           &            \\
    PoH$_2$   &   ---     & --290.25  & --261.88  &   --91.58  &  --91.56   &  --91.79  \\
    H$_2$O    &   ---     &  --93.62  &   --6.45  &    --6.45  &   --7.47   &   --6.73  \\
$\xi^\text{tot}$& --89.81 & --383.87  & --268.34  &   --98.03  &  --99.02   &  --98.52  \\
$\Delta \xi$  &     0.0   &   294.06  &   178.53  &      8.22  &     9.21   &     8.71  \\
%\bottomrule
\end{tabular}
\normalsize
\end{table}

All of the above let us conclude that, while FDE isotropic shieldings are
rather good and relatively much better as the systems become heavier, there are
still significant shortcomings in the description of the anisotropies in these systems.
The key to further improve the results is in ameliorating the leading FDE-LAO
contribution ($v$), and a possible way to do so is via the use of accurate
approximations to the non-additive kinetic energy
contributions.\cite{fux-jcp-132-164101-2010,artiukhin-jcp-142-234101-2015}

\subsection{Magnetizability}
\label{sub:results_magn}

The magnetizability tensors calculated with the DC Hamiltonian are summarized
in Table~\ref{tab:magnetizability}, where we present the isotropic
($\xi_\text{iso}$) and first anisotropy ($\xi_\text{aniso1}$) values. The results for the
second anisotropy, $\xi_\text{aniso2}$, can be found in the supplementary material.

The magnetizability is an extensive property and therefore in order to compare
to the supermolecular values one should obtain the tensors for both subsystems.
Another of its features, demonstrated by numerous studies, is that its value in
a molecule can be very well approximated by a sum of contributing atomic
susceptibilities,\cite{ruud-jacs-116-10135-1994,astrand-mrc-36-92-1998,bader-jcp-99-3683-1993}
with only few exceptions, e.g. aromatic
hydrocarbons,\cite{ruud-jacs-116-10135-1994} small molecules containing
fluorine\cite{ruud-jpca-105-9926-2001} or metal
clusters.\cite{schwerdtfeger-jcp-134-204102-2011} This curious additivity of
magnetizability tensor -- known as the Pascal's rule\cite{
pascal-annchimphys-19-5-1910, pascal-rsi-86-38-1948} --  has been attributed to
the local diamagnetic nature of atoms in
molecules\cite{flygare-chemrev-74-653-1974,ruud-jpca-105-9926-2001} and the
breakdown of this rule to the \emph{"long-range circulation of electrons"} not
accounted for in the atomic picture.\cite{flygare-chemrev-74-653-1974}

While Pascal's rule has been defined in terms of atomic susceptibilities, one
could consider the additivity of magnetizability of molecular assemblies in
terms of its constituent molecules. Recent studies\cite{kenneth-sem} 
have shown that Pascal's
rule is particularly useful when analysed in terms of the magnetically induced
current density, as the interaction of induced currents in neighbouring
molecules and the increase of paramagnetic component of magnetizability tensor
can be connected to the breakdown of the additivity rule.

\subsubsection{Isolated subsystems}
\label{subsub:results_magn_env}

As can be seen in Table~\ref{tab:magnetizability}, the $\Delta
\xi^\text{ME}_\text{iso}$ values are rather small in absolute (<0.1 for Se, --4.3 for Te
and --9.9 for Po, in SI units) as well as in relative terms. From these results,
one would be justified in considering that Pascal's rule holds rather well for
these systems.

The property shifts are larger for $\xi_\text{aniso1}$, with $\Delta
\xi_\text{aniso1}^\text{ME}$ that tend to increase with an increase of atomic number of
X (-5.5 for Se, 4.7 for Te and 8.7 for Po, in SI units) with the inverse trend
for $\Delta \xi_\text{aniso2}^\text{ME}$. More interesting, however, is that these
values amount overall to much larger relative differences (9.4\%, -5.8\%,
-9.7\% for Se, Te and Po for $\Delta \xi_\text{aniso1}^\text{ME}$, respectively), and
the change of sign along the series again indicates a complex interplay between
relativistic and environmental effects in this sort of embedding.

\subsubsection{Frozen Density Embedding}
\label{subsub:results_magn_fde}

For magnetizability, FDE-LAO contributions are also present in the property
gradient and in the expectation value part, thus it is again useful to consider
the relative importance of each of them. We shall here once again focus on
FDE(4) results, noting that FDE(0) results follow these closely but shows
slightly worse agreement with supermolecular values than FDE(4).  The $\Delta
\xi^\text{FDE(4)}_\text{iso}$ values calculated with no FDE-LAO terms are much larger in
magnitude than those for the isolated calculations. Unlike for $\sigma$, even
the inclusion of FDE-LAO$[v]$ terms yields $\Delta \xi^\text{FDE(4)}_\text{iso}$ values which are
significantly larger in magnitude (6.5 for Se, 12 for Te and 80.3 for Po, in
SI) than those for the isolated (``ME'') calculations.

The uncoupled kernel contribution ($v+w_u$) brings about little changes in
$\Delta \xi^\text{FDE(4)}_\text{iso}$ for Se but significant improvement for Te and,
especially, for Po.  As for the NMR shielding, the coupling kernel
($v+w_\text{all}$) term has a small effect, further leading to a decrease in
the $\Delta \xi^\text{FDE(4)}_\text{iso}$ values.

These results indicate that the additivity of $\xi_\text{iso}$ results from FDE
calculations with all second-order terms ($v+w_\text{all}$) is significantly
better than from the isolated (``ME'') calculations as the subsystem becomes heavier,
and may suggest an inflexion point between Se and Te where electronic effects
would become important enough for results to start deviating from Pascal's rule.

A similar trend is found for the $\Delta \xi^\text{FDE(4)}_\text{aniso1}$ values, with
FDE-LAO$[v]$ calculations underperforming isolated ones and the kernel
contributions being important for yielding good agreement with reference
values. As for the NMR shielding anisotropies, significant discrepancies with
respect to the supermolecular results remain. The performance of FDE for the
second anisotropy is slightly better but follows the same trends.

\subsection{NMR spin-spin coupling tensor}
\label{sub:results_spinspin}

The indirect reduced spin-spin coupling tensor calculated with the DC
Hamiltonian for the H--H$_b$, X--H and X--H$_b$ pairs of nuclei corresponding to
the XH$_2$ species are found in Table~\ref{tab:spinspin-dirac}.

\begin{table}
  \setlength\belowcaptionskip{5pt}
  \scriptsize
  \caption{
           \label{tab:spinspin-dirac}
           DC isotropic and anisotropic reduced indirect spin-spin couplings ($K_\text{iso}^\text{super}$ and $K_\text{aniso}^\text{super}$, in SI) for the H$_2$X subsystem in H$_2$X--H$_2$O, and absolute shifts ($\Delta K$, in SI) for the isolated (``ME'') and embedded (``FDE(4)'') H$_2$X molecules in the presence of H$_2$O.
          }
  \begin{tabular}{ll rrr r rrr}
  \toprule
%\multicolumn{2}{c}{\multirow{2}{*}{Nuclei}} &
\multicolumn{2}{c}{{Nuclei}} &
$K_\text{iso}^\text{super}$ &
$\Delta K_\text{iso}^\text{FDE(4)}$ &
$\Delta K_\text{iso}^\text{ME}$ &
&
$K_\text{aniso}^\text{super}$ &
$\Delta K_\text{aniso}^\text{FDE(4)}$ &
$\Delta K_\text{aniso}^\text{ME}$ \\
&&&&&&&&\\
\hline
&&&&&&&&\\
  H$_b$ &       Se      &      --11.22  & 1.52  &       7.25 && 113.79  &      --1.23&  1.25 \\
  H     &       Se      &      --16.04  & 0.07  &       1.51 && 110.68  &      --0.33& --1.01\\
  H$_b$ &       H       &      --0.77   & 0.00  &      --0.02&& 0.89    &       0.00 & --0.01\\
        &               &               &       &            &&         &            &       \\
  H$_b$ &       Te      &      --53.11  & 1.69  &       9.08 && 208.54  &      --0.42&  8.07 \\
  H     &       Te      &      --59.98  & 0.16  &       2.18 && 198.09  &      --0.44& --2.37\\
  H$_b$ &       H       &      --0.75   & 0.00  &      --0.03&& 0.42    &       0.00 &  0.00 \\
        &               &               &       &            &&         &            &       \\
  H$_b$ &       Po      &      --442.55 & 1.37  &      --1.25&& 429.17  &       3.10 &  41.96\\
  H     &       Po      &      --437.24 & 0.45  &       5.04 && 388.29  &      --0.18&  2.24 \\
  H$_b$ &       H       &      --0.61   & 0.00  &      --0.04&& 0.69    &       0.00 &  1.32 \\
\end{tabular}
\normalsize
\end{table}

\subsubsection{Isolated subsystems}
\label{subsub:results_spinspin_env}

The absolute values of isotropic one-bond spin-spin coupling constants (SSCCs)
involving the heavy nuclei increase significantly and, due to
relativity,\cite{autschbach-arnmrs-67-1-2009} for PoH$_2$ are around 30 times
larger to those in SeH$_2$. In relative terms, however, X--H SSCCs hardly change
for the isolated (``ME'') subsystems in relation to supermolecular values, with a slight
increase as the systems become heavier (from about 2 for Se or Te to 5
for Po, SI units), whereas for X--H$_b$ SSCCs the opposite trend is found (from 7--9
for Se/Te to --1 for Po, SI units).

The observation that environment effects are more important in the case of
Se--H$_b$ and Te--H$_b$ than of Se--H and Te--H, respectively,
is intuitive considering that most of the studied
one-bond spin-spin couplings are governed by Fermi contact
interactions,\cite{repisky-jackowski-jaszunski-book-2016} which probe the spin
density at the coupled nuclei - expected to be perturbed more on H$_b$ than on
the other H nucleus upon formation of hydrogen bond.  That said, the $\Delta
K^\text{ME}_\text{iso}$(Po--H$_b$) value is interesting in being the only one which is
negative and smaller than the $\Delta K^\text{ME}_\text{iso}$(Po--H) value, 
something that may indicate that the spin-spin coupling mechanism in
PoH$_2$ molecule is more complex and may be dominated by other interactions, for
instance spin-orbit-induced as in heavier interhalogen
diatomics,\cite{bryce-jacs-124-4894-2002} which may further be differently affected by
environmental effects.  Two-bond SSCCs (H--H$_b$) are in general very small for
all systems, though there is a small increase in absolute terms as the systems
become heavier.

For anisotropies, the values of $\Delta K^\text{ME}_\text{aniso}$ strongly increase for
X--H$_b$ as X becomes heavier, though it remains very small for X--H, and for
H--H$_b$ isolated (``ME'') results are largely the same as supermolecular ones.

\subsubsection{Frozen Density Embedding}
\label{subsub:results_spinspin_fde}

Unlike the other two properties discussed above, FDE contributions to the
response only enter here in the electronic Hessian, which greatly
simplifies the implementation.  Here we shall discuss SSCCs FDE(4) results
which are, like for magnetizabilities, better than FDE(0) results. We observe
that FDE performs better than mechanical embedding, with $\Delta K^\text{FDE(4)}_\text{iso}$ values
being consistently around 1--2 ppm for X--H$_b$ and smaller than 1 ppm for X--H or
H--H$_b$.

Similar trends as for $\Delta K^\text{FDE(4)}_\text{iso}$ are seen for $\Delta
K^\text{FDE(4)}_\text{aniso}$, with the latter being generally small and slightly negative
for all but $\Delta K^\text{FDE(4)}_\text{aniso}$(Po-H$_b$), which is of about 3 (SI units).

\subsection{A Comparison of DC and ZORA Hamiltonians}

Although SO-ZORA is known to yield rather different results from DC ones, it is
often sufficient in the determination of chemical shifts due to cancellation of
errors.\cite{malkin-jphyschemlett-4-459-2013,autschbach-mp-111-2544-2013}
Nevertheless when environmental effects are calculated as differences of absolute
shieldings, the cancellation of errors is not always guaranteed.

In recent years the differences in performance between the two Hamiltonians has
gained attention, with a number of studies reporting significant discrepancies
between SO-ZORA and DC values of NMR shielding tensors of heavy
nuclei,\cite{autschbach-mp-111-2544-2013,makulski-jms-1017-45-2012,malkin-jphyschemlett-4-459-2013,vicha-pccp-17-24944-2015}
which were explained by poor description of core orbitals of heavy elements by
ZORA Hamiltonian.\cite{autschbach-mp-111-2544-2013}

\begin{table}
  \scriptsize
  \caption{
           \label{tab:shieldings-adf}
          Absolute SO-ZORA isotropic and anisotropic shielding values ($\sigma_\text{iso}^\text{super}$ and $\sigma_\text{aniso}^\text{super}$, in ppm) of nuclei in H$_2$X (X = Se, Te, Po) subsystems in H$_2$X--H$_2$O, and absolute shifts ($\Delta \sigma$, in ppm) for the isolated (``ME'') and embedded (``FDE(4)'') H$_2$X molecules in the presence of H$_2$O.           
          }
  \begin{tabular}{l rrr r rrr}
  \toprule
Atom &
$\sigma_\text{iso}^\text{super}$ &
$\Delta \sigma_\text{iso}^\text{FDE(4)}$ &
$\Delta \sigma_\text{iso}^\text{ME}$ & 
 &
$\sigma_\text{iso}^{super}$ &
$\Delta \sigma_\text{iso}^\text{FDE(4)}$ &
$\Delta \sigma_\text{iso}^\text{ME}$ \\
&&&&&&\\
\hline
&&&&&&\\
Se	&2261.59&-11.30	 & 34.89 &&	628.15	& 4.43 &  4.89 \\
H$_b$	&30.10	&-0.77	 &--2.07 &&	23.95	& 5.83 &  7.22  \\
H	&32.55	&-0.03	 & 0.35	 &&	16.97	&--0.24&  0.11 \\
        &       &        &       &&             &      &       \\
Te	&4251.23&-9.61	 & 64.66 &&	1219.57	& 0.55 &  0.80 \\
H$_b$	&33.43	&-0.46	 &--1.59 &&	18.09	& 3.87 &  4.90  \\
H	&35.47	&0.00	 & 0.45	 &&	13.09	&--0.33 & --0.09  \\
        &       &        &       &&             &      &       \\
Po	&11168.82&-20.93 & 101.35&&	3138.04 &--45.36&--304.56 \\
H$_b$	&37.46	&-0.21	 &--0.82 &&	62.41	&--1.32 &-4.11  \\
H	&38.99	&0.00	 & 0.67	 &&	64.80	& 0.25 &-2.05  \\
%\bottomrule
\end{tabular}
\normalsize
\end{table}

It is therefore interesting to see how FDE and ME perform for the two
Hamiltonians. The results of our calculations of NMR shieldings and SSCCs with
SO-ZORA Hamiltonian are shown in Table~\ref{tab:shieldings-adf} and
Table~\ref{tab:spinspin-adf}, respectively.

A comparison of our SO-ZORA and DC results indicates that such error
cancellation occurs for SeH$_2$, since the results and trends of shieldings and
SSCCs are essentially the same for both Hamiltonians, for FDE and ME
calculations.
For TeH$_2$ both Hamiltonians yield largely similar results, but some
quantitative differences start to appear for $\Delta \sigma_\text{aniso}$ and rather
small differences for $\Delta \sigma_\text{iso}^\text{ME}$.  For PoH$_2$, on
the other hand, the differences are numerous: $\Delta \sigma_\text{iso}^\text{ME}$(Po)
for SO-ZORA already differs from DC value by 38 ppm, such difference for $\Delta
\sigma_\text{iso}^\text{FDE(4)}$ is about --16 ppm (DC using the
($v+w_\text{all}$) FDE-LAO terms), whereas for $\Delta
\sigma_\text{iso}^\text{ME}$(H$_b$) and $\Delta \sigma_\text{iso}^\text{FDE}$(H$_b$) these
discrepancies are of the order of a ppm. The differences between Hamiltonians are
also quite marked for anisotropies, where they amount to about 160 ppm for
$\Delta \sigma_\text{aniso}^\text{ME}$(Po) and 30 ppm for $\Delta
\sigma_\text{aniso}^\text{FDE}$(Po).

\begin{table}
  \setlength\belowcaptionskip{5pt}
  \scriptsize
  \caption{
           \label{tab:spinspin-adf}
           SO-ZORA isotropic and anisotropic reduced indirect spin-spin couplings ($K_\text{iso}^\text{super}$ and $K_\text{aniso}^\text{super}$, in SI) for the H$_2$X subsystem in H$_2$X--H$_2$O, and absolute shifts ($\Delta K$, in SI) for the isolated (``ME'') and embedded (``FDE(4)'') H$_2$X molecules in the presence of H$_2$O. 
          }
  \begin{tabular}{ll rrr r rrr}
  \toprule
%\multicolumn{2}{c}{\multirow{2}{*}{Nuclei}} &
\multicolumn{2}{c}{{Nuclei}} &
$K_\text{iso}^\text{super}$ &
$\Delta K_\text{iso}^\text{FDE(4)}$ &
$\Delta K_\text{iso}^\text{ME}$ &
&
$K_\text{aniso}^\text{super}$ &
$\Delta K_\text{aniso}^\text{FDE(4)}$ &
$\Delta K_\text{aniso}^\text{ME}$ \\
&&&&&&&&\\
\hline
&&&&&&&&\\
 H$_b$   &       Se      &      --12.63  & 1.08   &      6.58 && 130.61  &      --2.01&  1.15 \\
 H       &       Se      &      --16.41  & 0.87   &      1.88 && 127.57  &      --0.24& --1.03\\
 H$_b$   &       H       &      --1.11   &--0.03 &      --0.07&&--0.83   &       0.01 &  0.01 \\
         &               &               &       &            &&         &            &       \\
 H$_b$   &       Te      &      --55.69  & 1.84  &       9.82 && 226.45  &      --1.00&  8.95 \\
 H       &       Te      &      --64.05  &--0.02 &       1.44 && 214.87  &      --0.52& --2.61\\
 H$_b$   &       H       &      --0.94   &--0.02 &      --0.06&&--0.34   &       0.00 &  0.00 \\
         &               &               &       &            &&         &            &       \\
 H$_b$   &       Po      &      --439.33 & 1.02  &      --0.87&& 343.71  &       1.40 &  666.43 \\
 H       &       Po      &      --435.42 & 0.40  &       4.12 &&--315.34 &       0.98 &  7.19 \\
 H$_b$   &       H       &      --0.80   &--0.02 &      --0.07&& 0.78    &       0.00 &  0.08 \\
\end{tabular}
\normalsize
\end{table}

Large discrepancies between Hamiltonians are also seen for PoH$_2$ SSCCs, but
much more marked for anisotropies than for isotropic values (about a ppm for
FDE or ME calculations).

While our dataset is rather small for drawing more general conclusions, it
strengthens the case for a more thorough assessment of approximate Hamiltonians
such as ZORA for calculating NMR parameters of 6$p$ molecules.

\section{Conclusions and outlook}
\label{sec:conclusions}

In this paper we have described the implementation of frozen density embedding
contributions in a response theory framework, in combination with
four-component DC Hamiltonian to NMR indirect spin-spin couplings, NMR
shieldings, and magnetizabilities for mean-field approaches (DFT-in-DFT or
HF-in-DFT).

Due to the use of LAOs, which introduce the dependence of the electron- and
spin-density on the external magnetic field in the case of NMR shieldings and
magnetizabilities,  additional embedding contributions to the property gradient
(both properties) and expectation value (magnetizability only) arise, both for
the individual subsystems as well as introducing a coupling between these.

By performing DFT calculations on H$_2$X--H$_2$O (X = Se, Te, Po) model systems
we have been able to show the relative importance of these
additional contributions to the properties in question, while at the same time
confirming the findings of other studies that frozen density embedding is well suited
to the calculation of NMR indirect spin-spin couplings and NMR shieldings.

We have observed that the inclusion of the embedding potential in the FDE-LAO
property gradient contributions accounts for the bulk of the environment
effects, and that the heavier the center the more intra-subsystem FDE-LAO
kernel contributions are important for both NMR shieldings and
magnetizabilities, due to increasing importance of spin-density contributions.
Coupling kernel LAO contributions, by contrast, are in general rather small.

We have exploited the use of the magnetically induced currents to obtain NMR
shielding tensor via a real-space approach as well as to analyse, for the first
time, the differences between supermolecular and embedded calculations in
complement to the analysis of the electron density employed so far.  We
consider that the property density plots provide much clearer picture of where in
space the deficiencies in the FDE treatment manifest themselves compared to scalar
values of property shifts or unperturbed electron density plots as done prior to this work.

We present for the first time FDE contributions to magnetizabilities. Unlike
the case of the electric polarizability and in line with Pascal's rule, it
appears that one can reconstruct to rather good accuracy the tensor for the
supermolecular system from the tensors of the individual subsystems, obtained
without the FDE coupling terms in the response equations.  This may potentially
make FDE a more reliable route to obtaining molecular magnetizabilies than
other embedding approaches, since the whole system is treated
quantum-mechanically.

We have also compared our results to those obtained with the spin-orbit ZORA
Hamiltonian. Although the latter performs well, for the PoH$_2$--H$_2$O system we
have observed significantly different results between the Hamiltonians in the
description of environment effects on the isotropic NMR shieldings and
anisotropic spin-spin coupling constants.  This is in contrast with the common
expectation that relative shieldings are largely insensitive to changes in
Hamiltonians and therefore it would be worthwhile to verify whether that is
indeed the case for other systems containing the heaviest elements.

\section{Acknowledgements}

We would like to acknowledge inspiring discussions with Kenneth Ruud on the calculation of magnetizabilities, and with Christoph Jacob on the coupled response in FDE-NMR. We also acknowledge illuminating discussions with Michal Repisky on the question of magnetic balance in four-component calculations.

The members of the PhLAM laboratory acknowledge support from the CaPPA project (Chemical and Physical Properties of the Atmosphere), funded by the French National Research Agency (ANR) through the PIA (Programme d'Investissement d'Avenir) under contract ``ANR-11-LABX-0005-01'' as well as by the Ministry of Higher Education and Research, Hauts de France council and European Regional Development Fund (ERDF) through the Contrat de Projets Etat-Region (CPER) CLIMBIO (Changement climatique, dynamique de l'atmosph\`ere, impacts sur la biodiversit\'e et la sant\'e humaine).
Furthermore, ASPG acknowledges funding from the CNRS Institute of Physics (INP) via the PICS program (grant 6386), and computational time provided by the French national supercomputing facilities (grants DARI t2015081859, x2016081859).

\section{Appendix: Working expressions for FDE-LAO contributions}

In the DIRAC software, the quaternion algebra\cite{saue-jcp-111-6211-1999} is employed.
Thus, $\Omega_{pq;0}$ and $i\Omega_{pq,\mu}$ are calculated from the real and imaginary parts, respectively, of the generally quaternion overlap distribution matrix (Eq.~\ref{eq:Omega_k}), what allows to easily discern charge- and spin-density contributions to KS matrix and its derivatives.

\label{sec:appendix}
Here we present the working formulas for FDE-LAO contributions to the property gradient (Eq.~\ref{eq:fde_prop_grad1}-\ref{eq:fde_prop_grad3}) and to the expectation value part of the magnetizability tensor (Eq.~\ref{eq:d2_Eint_depsilonA_depsilonB_1}-\ref{eq:d2_Eint_depsilonA_depsilonB_2}) derived for closed-shell reference. They were obtained by first separating the number-density ($n$: $\rho_0 = -en$) and spin-density ($s=\sqrt{\rho_\mu \cdot \rho_\mu}$, $\mu \in \{x, y, z\}$) contributions, then by applying the local \emph{ansatz}, in which XC and kinetic energies are approximated by functions of local density variables:\cite{bast-ijqc-109-2091-2009}
  \begin{equation}
    E_\text{xck} = \int \epsilon_\text{xck}(Q^I \cup Q^{II}) d\vec{r},
    \label{eq:E_xck}
  \end{equation}
with $Q^M = \{n^M, s^M, (\nabla n \cdot \nabla n)^M, (\nabla n \cdot \nabla s)^M, (\nabla s \cdot \nabla s)^M \}$ for $M\in\{I,II\}$. This allows to express the FDE-LAO contributions to the property gradient of subsystem I (the expression for subsystem $II$ can be obtained by exchanging the labels $I$ and $II$) and to the expectation value part of the magnetizability tensor in terms of scalar and vector pre-factors ($a_0$, $\vec{b}_0$, $c_0^{M,N}$, $c_\mu^{M,N}$, $\vec{d}_0^{\,M,N}$, $\vec{d}_\mu^{\,M,N}$ for $M, N \in \{I, II\}$), summarized in Tables~\ref{tab:prefactors_scalar} and~\ref{tab:prefactors_vector}, while various perturbed densities are outlined in Table~\ref{tab:density_derivs} and discussed at length in literature.\cite{salek-cp-311-187-2005,bast-ijqc-109-2091-2009,olejniczak-jcp-136-014108-2012}

\clearpage
% \begin{strip}
  \begin{align}
  \begin{aligned}
    \frac{\partial}{\partial B_{\alpha}}
    \frac{\partial E_\text{int}}{\partial \kappa_{ai}^I}\biggr|_0
    &=
    -\int \left[
    a_0^I \breve{\Omega}_{ia;0}^{B_{\alpha};I} + \vec{b}_0^{\,I}\cdot\nabla \breve{\Omega}_{ia;0}^{B_{\alpha};I}
    \right]
    \text{d} \vec{r} \\
    &- \iint
    \left[
    c_0^{I,I} \Omega_{ia;0}^I + \vec{d}_0^{\,I,I}\cdot \nabla \Omega_{ia;0}^I
    +
    c_\mu^{I,I} \Omega_{ia;\mu}^I + \sum_{\mu=x,y,z} \vec{d}_\mu^{\,I,I}\cdot \nabla \Omega_{ia;\mu}^I
    \right]
    \text{d} \vec{r}_1\text{d} \vec{r}_2
    \\
    &-\iint
    \left[
    c_0^{I,II} \Omega_{ia;0}^I + \vec{d}_0^{\,I,II}\cdot \nabla \Omega_{ia;0}^I
    +
    c_\mu^{I,II} \Omega_{ia;\mu}^I + \sum_{\mu=x,y,z} \vec{d}_\mu^{\,I,II}\cdot \nabla \Omega_{ia;\mu}^I
    \right]
    \text{d} \vec{r}_1\text{d} \vec{r}_2
    \label{eq:fde_lao_prop_grad_appendix}
  \end{aligned}
  \end{align}
  \begin{align}
    \frac{\partial^2E_\text{int}}{\partial B_\alpha \partial B_\beta}
    \biggr|_0
   &= \sum_{M = I, II}
    \Big\{
    \int \left[
    a_0^M \breve{\Omega}_{ii;0}^{B_{\alpha}B_\beta;M} + \vec{b}_0^{\,M}\cdot\nabla \breve{\Omega}_{ii;0}^{B_{\alpha} B_\beta;M}
    \right]
    \text{d} \vec{r} \\
    &+ \iint
    \left[
    c_0^{M,M} \breve{\Omega}_{ii;0}^{B_\beta;M} + \vec{d}_0^{\,M,M}\cdot \nabla \breve{\Omega}_{ii;0}^{B_\beta;M}
    +
    c_\mu^{M,M} \breve{\Omega}_{ii;\mu}^{B_\beta;M} \right.
    \left. + \sum_{\mu=x,y,z} \vec{d}_\mu^{\,M,M}\cdot \nabla \breve{\Omega}_{ii;\mu}^{B_\beta;M}
    \right]
    \text{d} \vec{r}_1
    \text{d} \vec{r}_2
     \Big\}
    \nonumber \\
    &+\sum_{\substack{M,N = I, II \\ M \neq N}} \iint
    \left[
    c_0^{M,N} \breve{\Omega}_{ii;0}^{B_\beta;M} + \vec{d}_0^{\,M,N}\cdot \nabla \breve{\Omega}_{ii;0}^{B_\beta;M}
    \right.
    +
    \left.
    c_\mu^{M,N} \breve{\Omega}_{ii;\mu}^{B_\beta;M} + \sum_{\mu=x,y,z} \vec{d}_\mu^{\,M,N}\cdot \nabla \breve{\Omega}_{ii;\mu}^{B_\beta;M}
    \right]
    \text{d} \vec{r}_1
    \text{d} \vec{r}_2.
    \label{eq:fde_lao_expval_appendix}
  \end{align}
% \end{strip}

\begin{table}
  \caption{
           \label{tab:density_derivs}
           The derivatives of the general density component, $\rho_k$, in OMO basis. ${\cal P}_{\alpha\beta}$ denotes the permutation over indices $\alpha$ and $\beta$, $k \in \{0, x, y, z \}$. Basis functions labeled by $\mu$ and $\nu$ are centered on nuclei $K$ and $L$, respectively. Subsystem indices are skipped.
Second-order derivatives of overlap dsitribution involve the derivatives of LAO overlaps ($\Omega_{ii;k}^{B_\alpha}$ and $\Omega_{ii;k}^{B_\alpha B_\beta}$) and the brace notation of Helgaker and J\o rgensen\cite{helgaker-jcp-95-2595-1991} and are discussed in detailes elsewhere.\cite{paper}
          }
  {\renewcommand{\arraystretch}{2.5}
  \begin{tabular*}{\linewidth}{@{\extracolsep{\fill}}ll}
  \hline
      $\left. \frac{\partial \rho_k}{\partial \kappa_{pq}} \right|_{\kappa = 0} = -\Omega_{qp;k}$
      &
      $\left. \frac{\partial \rho_k}{\partial B_\alpha} \right|_{\vec{B} = 0} = \breve{\Omega}_{jj;k}^{B_\alpha}$
      \\
      $\left. \frac{\partial^2 \rho_k}{\partial \kappa_{pq} \partial B_\alpha} \right|_{\vec{B} = 0} = -\breve{\Omega}_{qp;k}^{B_\alpha}$
      &
      $\left. \frac{\partial^2 \rho_k}{\partial B_\alpha \partial B_\beta} \right|_{\vec{B} = 0} = \breve{\Omega}_{jj;k}^{B_\alpha B_\beta}$
      \\[1ex]
  \end{tabular*}
}
\offinterlineskip
  {\renewcommand{\arraystretch}{2.5}
  \begin{tabular*}{\linewidth}{@{}l@{}l@{}l}
      \hline
        $\breve{\Omega}_{pq;k}^{B_\alpha}$
      & $=$
      & $\frac{\text{ie}}{2}(\vec{R}_{KL}\times\vec{r})_{\alpha} c_{\mu p}^{*}c_{\nu q}\Omega_{\mu\nu;k} + \Omega_{\mu\nu;k}\left\{ c_{\mu t}^{*}T_{tp}^{B_\alpha*}c_{\nu q} + c_{\mu p}^{*}c_{\nu t}T_{tq}^{B_\alpha}\right\} $
      \\
        $\breve{\Omega}_{ii;k}^{B_\alpha B_\beta}$
     & $=$
     & $\Omega_{ii;k}^{B_\alpha B_\beta} + {\cal P}_{\alpha\beta} \{ T^\alpha, \Omega^{B_\beta} \}_{ii} + \{ T^{B_\alpha B_\beta}, \Omega \}_{ii} $
      \\
    & $+$
    & $\frac{1}{2} {\cal P}_{\alpha\beta} \left( \{ T^{B_\alpha}, \{T^{B_\beta}, \Omega\}\}_{ii} - \{ T^{B_\beta} T^{B_\alpha}, \Omega \}_{ii;k} \right)$
      \\[1ex]
\end{tabular*}
}
\end{table}

\begin{table}
  \caption{
           \label{tab:prefactors_scalar}
           Scalar prefactors derived for closed-shell reference. $a_0^I = v_{emb;0}^I$
          }
  {\renewcommand{\arraystretch}{2.5}
  \begin{tabular*}{\linewidth}{lcl}
  \hline
     $\vec{b}_0^{\,I}$
   & $=$
   & $2 \left(\frac{\partial \varepsilon_{xck}}{\partial (\nabla n \cdot \nabla n)}\biggr|_{tot} \nabla n^{tot} - \frac{\partial \varepsilon_{xck}}{\partial (\nabla n \cdot \nabla n)}\biggr|_{I} \nabla n^{I} \right)$
     \\
     $c_0^{I,I}$
   & $=$
   & $\left( \frac{\partial^2 \varepsilon_{xck}}{\partial n^2}\biggr|_{tot} - \frac{\partial^2 \varepsilon_{xck} }{\partial n^2}\biggr|_{I} \right) \breve{\Omega}_{jj;0}^{I;B_\alpha}$\\
 \multicolumn{3}{l}{$  + 2\left( \frac{\partial^2 \varepsilon_{xck}}{\partial n \partial (\nabla n \cdot \nabla n)}\biggr|_{tot} \cdot \nabla n^{tot} - \frac{\partial^2 \varepsilon_{xck} }{\partial n \partial (\nabla n \cdot \nabla n)}\biggr|_{I}\cdot \nabla n^I  \right)\cdot \nabla \breve{\Omega}_{jj;0}^{I;B_\alpha} $}
     \\
     $\vec{d}_0^{\,I,I}$
   & $=$
   & $2\left( \frac{\partial^2 \varepsilon_{xck}}{\partial n \partial (\nabla n \cdot \nabla n)}\biggr|_{tot} \cdot \nabla n^{tot} - \frac{\partial^2 \varepsilon_{xck} }{\partial n \partial (\nabla n \cdot \nabla n)}\biggr|_{I}\cdot \nabla n^I  \right)\cdot  \breve{\Omega}_{jj;0}^{I;B_\alpha}$\\
   \multicolumn{3}{l}{$+4\left( \frac{\partial^2 \varepsilon_{xck}}{\partial (\nabla n \cdot \nabla n)^2}\biggr|_{tot} \cdot \nabla n^{tot}\cdot\nabla n^{tot} - \frac{\partial^2 \varepsilon_{xck} }{\partial (\nabla n \cdot \nabla n)^2}\biggr|_{I}\cdot \nabla n^I\cdot\nabla n^I  \right)\cdot \nabla \breve{\Omega}_{jj;0}^{I;B_\alpha}$}\\
   \multicolumn{3}{l}{$+2\left( \frac{\partial \varepsilon_{xck}}{\partial (\nabla n \cdot \nabla n)}\biggr|_{tot}  - \frac{\partial \varepsilon_{xck} }{\partial (\nabla n \cdot \nabla n)}\biggr|_{I} \right)\cdot \nabla \breve{\Omega}_{jj;0}^{I;B_\alpha}$}
     \\
     $c_0^{I,II}$
   & $=$
   & $ \frac{\partial^2 \varepsilon_{xck}}{\partial n^2}\biggr|_{tot}\breve{\Omega}_{jj;0}^{II;B_\alpha} + \int \frac{1}{|\vec{r}-\vec{r}_1|}\breve{\Omega}_{jj;0}^{II;B_\alpha}(\vec{r}_1) \text{d}\vec{r}_1 $ \\
   \multicolumn{3}{l}{$+2 \frac{\partial^2 \varepsilon_{xck}}{\partial n \partial (\nabla n \cdot \nabla n)}\biggr|_{tot} \cdot \nabla n^{tot} \cdot \nabla \breve{\Omega}_{jj;0}^{II;B_\alpha}$} \\
     $\vec{d}_0^{\,I,II}$
   & $=$
   & $2 \frac{\partial^2 \varepsilon_{xck}}{\partial n \partial (\nabla n \cdot \nabla n)}\biggr|_{tot} \cdot \nabla n^{tot} \cdot \breve{\Omega}_{jj;0}^{II;B_\alpha}$\\
   \multicolumn{3}{l}{$+4 \frac{\partial^2 \varepsilon_{xck}}{\partial (\nabla n \cdot \nabla n)^2}\biggr|_{tot} \cdot \nabla n^{tot}\cdot\nabla n^{tot} \cdot \nabla \breve{\Omega}_{jj;0}^{II;B_\alpha}
   +2 \frac{\partial \varepsilon_{xck}}{\partial (\nabla n \cdot \nabla n)}\biggr|_{tot} \cdot \nabla \breve{\Omega}_{jj;0}^{II;B_\alpha}$} \\
\end{tabular*}
}
\end{table}

\begin{table}
  \caption{
           \label{tab:prefactors_vector}
           Vector prefactors derived for closed-shell reference.
          }
  {\renewcommand{\arraystretch}{2.5}
  \begin{tabular*}{\linewidth}{lcl}
  \hline
     $c_\mu^{I,I}$
   & $=$
   & $\left( \frac{\partial^2 \varepsilon_{xck}}{\partial s^2}\biggr|_{tot} - \frac{\partial^2 \varepsilon_{xck} }{\partial s^2}\biggr|_{I} \right) \breve{\Omega}_{jj;\mu}^{I;B_\alpha}$ \\
\multicolumn{3}{l}{$+\left( \frac{\partial^2 \varepsilon_{xck}}{\partial s \partial (\nabla n \cdot \nabla s)}\biggr|_{tot} \cdot \nabla n^{tot} - \frac{\partial^2 \varepsilon_{xck} }{\partial s \partial (\nabla n \cdot \nabla s)}\biggr|_{I}\cdot \nabla n^I  \right)\cdot \nabla \breve{\Omega}_{jj;\mu}^{I;B_\alpha}$} \\
     $\vec{d}_\mu^{\,I,I}$
   & $=$
   & $\left( \frac{\partial^2 \varepsilon_{xck}}{\partial s \partial (\nabla n \cdot \nabla s)}\biggr|_{tot} \cdot \nabla n^{tot} - \frac{\partial^2 \varepsilon_{xck} }{\partial s \partial (\nabla n \cdot \nabla s)}\biggr|_{I}\cdot \nabla n^I  \right)\cdot  \breve{\Omega}_{jj;\mu}^{I;B_\alpha}$\\
\multicolumn{3}{l}{$ +\left( \frac{\partial^2 \varepsilon_{xck}}{\partial (\nabla n \cdot \nabla s)^2}\biggr|_{tot} \cdot \nabla n^{tot} \cdot \nabla n^{tot} - \frac{\partial^2 \varepsilon_{xck} }{\partial (\nabla n \cdot \nabla s)^2}\biggr|_{I}\cdot \nabla n^I \cdot \nabla n^I  \right) \cdot \nabla \breve{\Omega}_{jj;\mu}^{I;B_\alpha}$}\\
\multicolumn{3}{l}{$+2\left( \frac{\partial \varepsilon_{xck}}{\partial (\nabla s \cdot \nabla s)}\biggr|_{tot}  - \frac{\partial \varepsilon_{xck}}{\partial (\nabla s \cdot \nabla s)}\biggr|_{I} \right)\cdot \nabla \breve{\Omega}_{jj;\mu}^{I;B_\alpha}$} \\
     $c_\mu^{I,II}$
   & $=$
   & $\frac{\partial^2 \varepsilon_{xck}}{\partial s^2}\biggr|_{tot} \breve{\Omega}_{jj;\mu}^{II;B_\alpha}
   +\frac{\partial^2 \varepsilon_{xck}}{\partial s \partial (\nabla n \cdot \nabla s)}\biggr|_{tot} \cdot \nabla n^{tot} \cdot \nabla \breve{\Omega}_{jj;\mu}^{II;B_\alpha}$
     \\
     $\vec{d}_\mu^{\,I,II}$
   & $=$
   & $\frac{\partial^2 \varepsilon_{xck}}{\partial s \partial (\nabla n \cdot \nabla s)}\biggr|_{tot} \cdot \nabla n^{tot} \cdot  \breve{\Omega}_{jj;\mu}^{II;B_\alpha}$\\
   \multicolumn{3}{l}{$+\frac{\partial^2 \varepsilon_{xck}}{\partial (\nabla n \cdot \nabla s)^2}\biggr|_{tot} \cdot \nabla n^{tot} \cdot \nabla n^{tot}  \cdot \nabla \breve{\Omega}_{jj;\mu}^{II;B_\alpha}
   +2 \frac{\partial \varepsilon_{xck}}{\partial (\nabla s \cdot \nabla s)}\biggr|_{tot} \cdot \nabla \breve{\Omega}_{jj;\mu}^{II;B_\alpha}$}
      \\[1ex]
\end{tabular*}
}
\end{table}

%\bibliography{manuscript}
%\bibliographystyle{apsauth4-1}
%\bibliographystyle{plain}
%\bibliographystyle{unsrtnat}
%\bibliographystyle{unsrt}

\end{document}